\documentclass[floats,floatfix,amssymb,prd,twocolumn,superscriptaddress,nofootinbib]{revtex4-1}

\usepackage{subcaption}
\usepackage{ragged2e}
\DeclareCaptionJustification{justified}{\justifying}
\makeatletter
\newcommand{\subsetsim}{\mathrel{\mathpalette\subset@sim\relax}}
\newcommand{\subset@sim}[2]{%
  \vtop{\offinterlineskip\m@th
    \ialign{\hfil##\cr
      $#1\subset$\cr\noalign{\kern0.5pt}\scalebox{0.9}{$#1\sim$}\cr
    }%
  }%
}
\makeatother

\usepackage{amssymb,amsmath,verbatim,mathtools,needspace,enumitem,etoolbox,graphicx,physics,microtype,afterpage,bm}
\usepackage[dvipsnames, usenames]{xcolor}
\definecolor{linkcolor}{rgb}{0.0,0.3,0.5}
\usepackage{booktabs}
\usepackage[unicode, colorlinks=true, linkcolor=linkcolor, citecolor=linkcolor, filecolor=linkcolor,urlcolor=linkcolor, pdfusetitle]{hyperref}
\usepackage[all]{hypcap}
\usepackage[T1]{fontenc}
\usepackage[utf8]{inputenc}
\usepackage{tabularx}
\usepackage{float}
\renewcommand{\arraystretch}{1.4}

\usepackage{multirow}
\usepackage{pifont}
\usepackage{lmodern}

\usepackage{multirow}

\allowdisplaybreaks
\usepackage{tikz}
\usepackage{color}
\usepackage{framed}
\usepackage{hyperref}
\hypersetup{colorlinks, citecolor=bluscuro, linkcolor=black, urlcolor=bluscuro}
\definecolor{rossos}{cmyk}{0,1,1,0.55}
\definecolor{bluscuro}{rgb}{0.15, 0.2, .85}
\definecolor{bluchiaro}{cmyk}{1,.3,0.,0.1}
\definecolor{ForestGreen}{rgb}{0.13, 0.55, 0.13}

\newcommand{\apjl}{"Astrophys. J. Lett."}

\newcommand{\aapr}{"Astronomy and Astrophysics Reviews"}

\newcommand{\mnras}{"Mon. Not. Roy. Astron. Soc."}

\def\f{\frac}

\newcommand{\chimax}{\chi_\text{\tiny \rm max}}
\def\chieff{\chi_\text{\tiny \rm eff}}

\def\bea{\begin{eqnarray}}
\def\eea{\end{eqnarray}}

\def\co{{\text{\tiny \rm cut-off}}}
\def\d{{\mathrm{d}}}
\def\PBH{\text{\tiny{PBH}}}

\newcommand{\bs}{\begin{subequations}}
\newcommand{\es}{\end{subequations}}

\newcommand{\be}{\begin{equation}}
\newcommand{\ee}{\end{equation}}
\renewcommand{\d}{{\rm d}}

\newcommand{\llp}{\left [}
\newcommand{\rrp}{\right ]}

\def\lsim{\mathrel{\rlap{\lower4pt\hbox{\hskip0.5pt$\sim$}}
    \raise1pt\hbox{$<$}}}         
\def\gsim{\mathrel{\rlap{\lower4pt\hbox{\hskip0.5pt$\sim$}}
    \raise1pt\hbox{$>$}}}         

\newcommand{\spp}{\hspace{.19 cm}}

\newcommand{\sppdd}{\hspace{.01 cm}}

\newcommand{\sapienza}{Dipartimento di Fisica, Sapienza Università 
	di Roma, Piazzale Aldo Moro 5, 00185, Roma, Italy}
\newcommand{\infn}{INFN, Sezione di Roma, Piazzale Aldo Moro 2, 00185, Roma, Italy}

\begin{document}

\title{Searching for 
mass-spin correlations in the population of
\\
gravitational-wave events: 
the GWTC-3 case study}

\author{Gabriele Franciolini}
\email{gabriele.franciolini@uniroma1.it}
\affiliation{\sapienza}
\affiliation{\infn}

\author{Paolo Pani}
\affiliation{\sapienza}
\affiliation{\infn}


\begin{abstract}
One fundamental goal of the newly born gravitational wave astronomy is discovering the origin of the observed binary black hole mergers. 
Towards this end, identifying features in the growing wealth of data may help in distinguishing different formation pathways. 
While large uncertainties still affect the binary formation models, spin-mass relations remain characteristic features of specific classes of channels.  
By focusing on the effective inspiral spin $\chi_\text{\tiny eff}$, the best reconstructed spin-related merger parameter, we show that current GWTC-3 data support the hypothesis that a fraction of events may display mass-spin correlations 
similar to one expected by dynamical formation channels of either
astrophysical or primordial nature.
We quantify the Bayesian evidence in favour of those models, which are substantially preferred when compared to the Gaussian phenomenological model adopted to describe the distribution of
$\chi_\text{\tiny eff}$ in the recent LIGO/Virgo/KAGRA population analyses. 
\end{abstract}

\maketitle

\section{Introduction}

The third gravitational-wave~(GW) catalogue~\cite{LIGOScientific:2021djp} recently released by the LIGO/Virgo/KAGRA collaboration~(LVKC)~\cite{TheLIGOScientific:2014jea,TheVirgo:2014hva}
increased the number of GW detections to around $90$ events --~most of which are binary black holes~(BBHs)~-- and almost doubled the number of BBH events previously available~\cite{Abbott:2020niy,Abbott:2020gyp}.
Following the ever-growing number of detections, we can characterize with increasing accuracy the properties of the underlying BH population. 
Various astrophysical formation scenarios were proposed to explain the BBH formation~\cite{Mandel:2018hfr,Mapelli:2018uds,Mandel:2021smh,Broekgaarden:2021efa}. 
In particular, the observed excess of massive BHs (around approximately $35 M_\odot$) may be due to hierarchical mergers of smaller objects~\cite{Gerosa:2017kvu,Fishbach:2017dwv,Rodriguez:2019huv,Gerosa:2019zmo,Gerosa:2021hsc} or the pile up from the evolution of massive stars just below the pair-instability supernova mass gap~\cite{Heger:2001cd,Belczynski:2016jno,Gayathri:2021xwb}.
Another interesting possibility is that a subpopulation of primordial BHs~(PBHs)~\cite{Sasaki:2018dmp,Green:2020jor,Franciolini:2021nvv} may be contributing to the observed GW events~\cite{Franciolini:2021tla}, while being a compelling explanation for the detected mass-gap events \cite{Clesse:2020ghq,DeLuca:2020sae}
 (such as GW190814~\cite{LIGOScientific:2020zkf} appearing in the hypothesized low mass gap between neutron stars and BHs, and GW190521~\cite{LIGOScientific:2020iuh,LIGOScientific:2020ufj} above a theorized pair instability
limit), which might be challenging to explain in the standard astrophysical scenarios.

A possible subpopulation of PBHs is particularly intriguing as it would provide a portal linking GW observations to early-universe cosmology and particle physics. PBHs can be produced during the radiation-dominated era 
from the collapse of very large inhomogeneities~\cite{Zeldovich:1967lct,Hawking:1974rv,Chapline:1975ojl,Carr:1975qj} across a wide mass range~\cite{Ivanov:1994pa,GarciaBellido:1996qt,Ivanov:1997ia,Blinnikov:2016bxu}. Although various constraints were set on the abundance of these objects (see~\cite{Carr:2020gox} for a review), in certain mass ranges PBHs could comprise the entirety of the dark matter and be the seeds of supermassive BHs at high redshift~\cite{2010A&ARv..18..279V,Clesse:2015wea,Serpico:2020ehh}.
Furthermore, it was shown that PBHs could contribute to a fraction of the BH merger events detected so far~\cite{Bird:2016dcv,Sasaki:2016jop,Eroshenko:2016hmn, Wang:2016ana, Ali-Haimoud:2017rtz,Chen:2018czv,Raidal:2018bbj, Hutsi:2019hlw, Vaskonen:2019jpv, Gow:2019pok,Wu:2020drm,Jedamzik:2020ypm,Jedamzik:2020omx,DeLuca:2020qqa,Bhagwat:2020bzh,Hall:2020daa,DeLuca:2020jug,Wong:2020yig,Hutsi:2020sol,Kritos:2020wcl,Deng:2021ezy,Kimura:2021sqz,Bavera:2021wmw,Chen:2021nxo,Franciolini:2021tla} and of those detectable by future third-generation facilities~\cite{DeLuca:2021wjr,DeLuca:2021hde,Pujolas:2021yaw,Ng:2021sqn}.

In order to derive information about the intrinsic population of BBHs, the LVKC used phenomenological models built with the aim of capturing the properties of the mass, spin, and redshift distribution of BBHs~\cite{Abbott:2020gyp,2021arXiv211103634T}.
Those analyses highlight key characteristics that BBH models should reproduce in order to be compatible with current data and may guide current understanding of binary formation channels. 
Therefore, identifying new features in the growing wealth of data may provide novel insights on which mechanism was responsible for the generation of the observed population of binaries, or provide evidence for multiple populations generated by different channels~\cite{Zevin:2020gbd,Wong:2020ise,Hutsi:2020sol,Franciolini:2021tla,Li:2022jge}.

In this work we focus on the BH spins in binaries and, in particular, their correlation with the masses. 
At the leading post-Newtonian order, the inspiral GW signals depend only on a single effective spin parameter~\cite{PhysRevD.64.124013,Racine:2008qv,Ajith:2009bn}
\begin{equation}\label{eq:chieff}
    \chi_\text{\tiny eff} 
    \equiv 
    \frac{
    \chi_1 \cos{\alpha_1} + q \chi_2 \cos{\alpha_2} 
    }{1+q},
\end{equation} 
which is a function of the mass ratio $q=m_2/m_1\leq1$, of both BH spin magnitudes 
$\chi_i=J_i/m_i^2$ (with $0\leq \chi_i\leq 1$), and of their orientation with respect to the orbital angular momentum, parametrized by the tilt angles $\alpha_i$.
Even though the chirp mass 
can be constrained much better than $\chi_\text{\tiny eff}$, mass distributions expected in astrophysical or primordial models remain uncertain (or,
in the latter case, completely unconstrained a priori). 
On the contrary, mass-spin relations may provide distinctive features of particular BBH models. 
For example, in the most common formation scenario PBHs are formed with negligible natal spin~\cite{DeLuca:2019buf,Mirbabayi:2019uph}, and PBH binaries are characterised by a mass-spin correlation solely induced by accretion effects~\cite{DeLuca:2020qqa}. In the astrophysical dynamical formation scenario, instead, the existence of multi-generation mergers implies well characterised mass-spin correlations~\cite{Gerosa:2021mno}. 
Typically, in the literature, the distinction between astrophysical binaries born in isolation or assembled dynamically is only based on the nature of $\chieff$ being restricted to positive values or distributed symmetrically around zero~\cite{PhysRevD.96.023012,Farr:2017uvj,Farr:2017gtv,Tiwari:2018qch,PhysRevD.100.043012,Safarzadeh:2020mlb,Stevenson:2017dlk,Roulet:2018jbe,Miller:2020zox}. In this work we try to go beyond this simple requirement and 
 devise a refined dynamical model which includes the correlation of $\chieff$ with the masses.

Interestingly, the LVKC population analysis found that $q$ and $\chieff$ are correlated, 
while the distribution of the absolute value of the spin projection along the $z$-axis (i.e., the one orthogonal to the orbital plane) is allowed to broaden above roughly  $30M_\odot$ \cite{2021arXiv211103634T}. While the former property appears to be in contrast with what is expected from BBH population models~\cite{Callister:2021fpo} (with possibly the exception of few special cases of isolated evolution~\cite{Bavera:2020uch} and formation in AGN disks~\cite{Mckernan:2017ssq,Stone:2016wzz,Tagawa:2019osr}),
the latter may hint to a correlation similar to the one expected in dynamical formation scenarios (either of primordial or astrophysical origin) lurking in the data (see also Refs.~\cite{Hoy:2021rfv,Wang:2021clu}).
In this work, we search for mass-spin correlations motivated by both dynamical and primordial formation channels, showing that both models are statistically preferred when compared to a Gaussian phenomenological distribution of $\chieff$ as the one adopted in~\cite{2021arXiv211103634T}.

\section{BBH models and effective spin distributions} 

In this section, we review the BBH models we adopt in this work and their predictions for the distribution of $\chieff$.

\subsection{Gaussian effective spin phenomenological model}

In the recent population analyses of the GWTC-3 data, the LVKC has adopted various phenomenological spin models~\cite{2021arXiv211103634T}. 
 
The {\sc Gaussian} model (hereafter often denoted by G) assumes that $\chi_\text{\tiny eff}$ follows a normal distribution with mean $\mu$ and variance $\sigma$, normalised within the physical range $[-1,1]$, as~\cite{Roulet:2018jbe,Miller:2020zox}
\begin{align}
    p_\text{\tiny pop}^{\text{\tiny G}}(\chi_\text{\tiny eff} | \mu, \sigma)
    = {\cal N}(\mu,\sigma) \exp\llp {-{(\chi_\text{\tiny eff} - \mu )^2}/{2 \sigma^2}}\rrp.
\end{align}
The explicit expression of the normalisation ${\cal N}(\mu,\sigma)$ can be found, for example, in Eq.~(4) of Ref.~\cite{Miller:2020zox}.

Recent analyses of the LVKC data~\cite{Callister:2021fpo,2021arXiv211103634T}
also reported a preference for an extended \textsc{Gaussian} model (hereafter often denoted by G$_\text{\tiny corr}$) with mean and standard deviation evolving with the mass ratio as~\cite{Callister:2021fpo}
   \begin{align}
    \mu(q|\mu_{0},\alpha) 
    &= \mu_{0} + \alpha (q-1)   ,
    \nonumber \\
\log_{10}\sigma(q|\sigma_{0},\beta) 
&= \log_{10}\sigma_{0} + \beta(q-1),
\label{eqcorrq}
   \end{align}
for some parameters $\sigma_0$, $\mu_0$, $\alpha$, and $\beta$.
In particular, Ref.~\cite{Callister:2021fpo} found
a preference for asymmetric binaries ($q < 1$) to have positive $\chieff$ (i.e. $\alpha<0$). As we shall discuss, 
this trend is different from the one seen in dynamical formation scenarios (either of astrophysical or primordial origin) in which $\chieff$ is always symmetric around zero while only a larger {\it width} of the distribution may correlate with smaller $q$. 

We will mix dynamical formation channels of both astrophysical and primordial origin with the above \textsc{Gaussian} benchmark models, both 
including and excluding the $q-\chieff$ correlation, and show that this extension of the \textsc{Gaussian} model does not affect our conclusions.

\begin{figure*}[t]
\centering
\includegraphics[width=0.4935\textwidth]{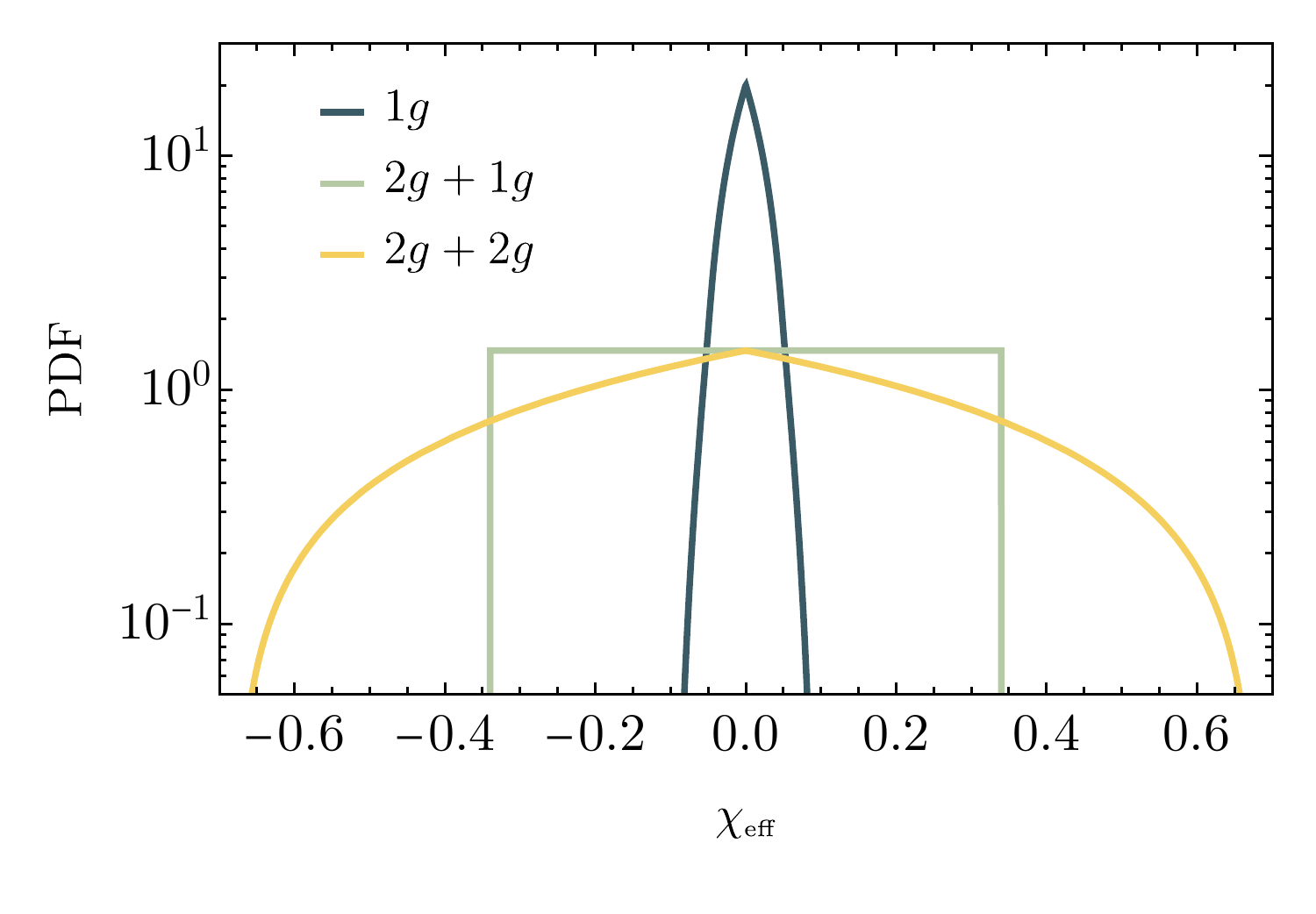}
\includegraphics[width=0.4865\textwidth]{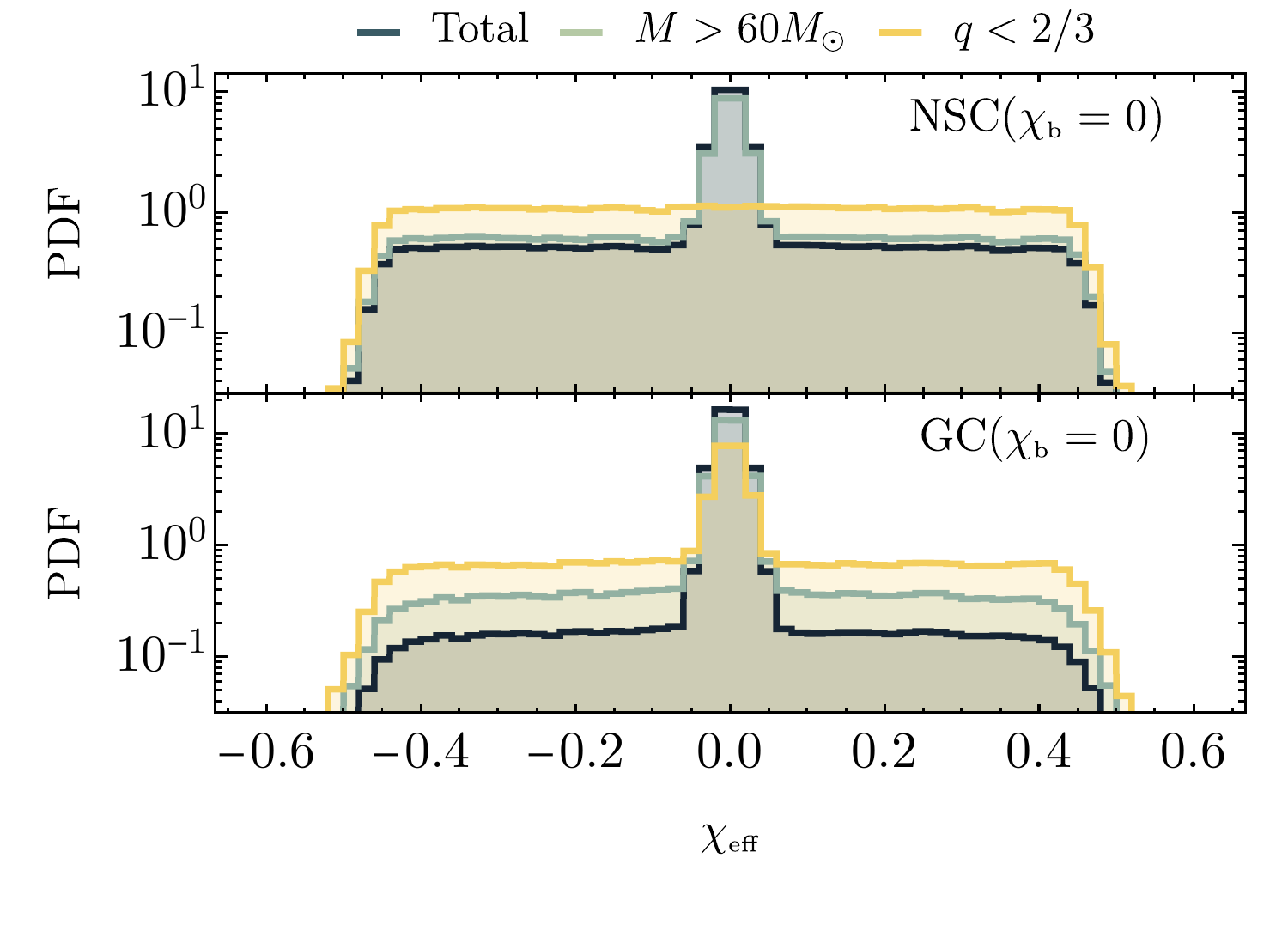}
\caption{ 
{\bf Left:} Effective distribution of $\chieff$ for $1g$, $2g+1g$ and $2g+2g$ mergers, assuming $\chi_\text{\tiny max} = 0.1$ and a distribution of mass ratio sharply peaked towards unity.  
{\bf Right:}
Distribution of $\chieff$ in the population synthesis models GC and NSC as defined in Ref.~\cite{Zevin:2020gbd}. 
The dark line indicates the total distribution while the green (yellow) lines correspond to the distribution for massive (asymmetric) sources, highlighting the mass dependence of the mixing fraction between the various generations. 
A realistic distribution of $q$ widens the $2g+1g$ distribution and generates a milder drop-off in the tails with respect to Eq.~\eqref{2g+1gder}, while the fraction of $2g+2g$ mergers is smaller and falls below the scale of the plot. 
We point out that the NSC model does not present a sharp peak at $\chieff \simeq 0$ for $q<2/3$ due to the pairing function assumed in that model, which produces predominantly symmetric (i.e. $q\simeq 1$) first-generation mergers. 
This does not apply to the GC model, which assumes a broader distribution of mass ratio already for first-generation mergers
(see Ref.~\cite{Zevin:2020gbd} and references therein for more details).
} 
\label{fig:ABHdyn}
\end{figure*}

\subsection{Astrophysical dynamical BH models}\label{sec:th_mspins_Astro}

In general, astrophysical models may give rise to BBHs either through isolated evolution of stellar binaries in the galactic field or dynamical formation in star clusters~\cite{Mandel:2018hfr,Mapelli:2018uds,Mandel:2021smh,Broekgaarden:2021efa}.

Field formation channels are predicted to produce preferentially aligned spin orientations, which translate in a distribution of $\chieff$ which is centered at positive values with potentially larger $\chieff$ for lighter binaries (see e.g.~\cite{Belczynski:2017gds,Qin:2018vaa,Gerosa:2018wbw,Bavera:2020inc,Bavera:2020uch}). This general prediction may be compatible with the dominant fraction of events in the catalog and it is in sharp contrast with the prediction of dynamical channels producing random spin orientations and a symmetric distribution of $\chieff$. In the rest of this paper, we will assume that the population of field binaries is described by the aforementioned \textsc{Gaussian} models and will only focus on the mass-spin correlations expected in the dynamical channels. For simplicity, we will also refer to the dynamical astrophysical model simply as ABH model.

The phenomenological model we adopt for describing the distribution of $\chieff$ is built as an extension of the one introduced in Ref.~\cite{Baibhav:2020xdf} (see their Appendix~A), and validated comparing it with state-of-the-art population models, as discussed in the following. 
As the dynamical formation scenarios give rise to different $\chieff$ distributions for binaries of various generations (see Fig.~\ref{fig:ABHdyn}), we analytically describe them separately under simplifying assumptions and present a simple educated phenomenological model of  their differential contribution as a function of total mass $M$ and mass ratio $q$. 

\noindent {\bf \em First generation (1g).}
The BBH assembled dynamically are characterised by random and independent spin orientations. As a consequence, the distribution of $\chieff$ only depends on the spin distribution which we assume, following Ref.~\cite{Baibhav:2020xdf}, to be uniform, i.e. $\chi_i  \in {\rm U} [0,\chi_\text{\tiny max}]$.
Averaging over the possible angles $\alpha_i$, one obtains the distribution $p_\text{\tiny pop}^{1g} (\chieff | \chi_\text{\tiny max})$ whose explicit expression can be found in Appendix~A of Ref.~\cite{Baibhav:2020xdf}, see their Eq.~(A14).

\noindent 
{\bf \em Second-generation (2g+1g)}
Second generation BHs are those born out of a previous merger. Working under the assumption of small-$\chimax$, the spin of $1g$ BHs is neglibigle, while $2g$ remnants are characterised by a final spin $\chi_f\simeq0.68$ (see e.g. Ref.~\cite{Hofmann:2016yih}). Furthermore, the mass ratio is expected to be smaller than unity.
Therefore, in Eq.~\eqref{eq:chieff} the term $q\chi_2$ is much smaller than $\chi_1=\chi_f$ and
\be
\chieff \simeq \f{\chi_f \cos\alpha_1}{1+q}\,.
\ee
Since $\cos\alpha_1$ is distributed uniformly in ${\rm U}[-1,\,1]$, the distribution of $\chieff$ depends on the pairing function describing the mass ratio distribution of second generation binaries, whose exact details depend on the specific model. 
The resulting distribution can be approximated as~\cite{Baibhav:2020xdf}
\be\label{2g+1gder}
p_\text{\tiny pop}^{2g+1g}(\chieff) \simeq \f{1}{\chi_f} 
\quad \text{for} \quad
|\chieff|\le \f{\chi_f}{2}\,,
\ee
and is displayed in the left panel of Fig.~\ref{fig:ABHdyn}.
We will show in the following how (a slightly modified version of) this functional form approximates quite well state-of-the-art dynamical models. 

\noindent
{\bf \em Multi-generation (2g+2g)}
Finally, following similar arguments, for $2g+2g$ mergers one finds the approximated distribution~\cite{Baibhav:2020xdf}
\be
p_\text{\tiny pop}^{2g+2g}(\chieff) \simeq \f{1}{\chi_f} \left(1 \!-\! \f{\vert\chieff\rvert}{\chi_f}\right)
\quad
\text{for}
\quad 
|\chieff|\le\chi_f\,.
\ee

We compare the expectation of such an effective description of the $\chieff$ distribution to the state-of-the-art models describing the dynamical formation of binaries in Globular Clusters~(GCs) and Nuclear Star Clusters~(NSCs) as released in Ref.~\cite{Zevin:2020gbd} based on the studies in Refs.~\cite{Antonini:2018auk,Rodriguez:2019huv} and references therein.
For both models, we consider the initial assumption of vanishing spins at birth (i.e. $\chi_\text{\tiny b} = 0$ in the notation of~\cite{Zevin:2020gbd}), which is found to best fit the observed population of GW events~\cite{Zevin:2020gbd,Franciolini:2021tla}.
As we can see in the right panel of Fig.~\ref{fig:ABHdyn}, 
the distribution of $\chieff$ in both the GC and NSC models is simply built out of a dominant contribution from the $1g$ BBHs with small $\chi_\text{\tiny max}$ and a plateau comprised of $2g+1g$ mergers, with a fraction that depends on both the total mass $M$ and the mass ratio $q$. 
As we can see by comparing the right and left panels of Fig.~\ref{fig:ABHdyn}, the simplified analytical $2g+1g$ model captures the main features of the synthetic population, but with a slightly smaller support.
In order to capture the realistic distribution of $q$ not accounted for in the simplified $2g+1g$ model, we modify Eq.~\eqref{2g+1gder} and assume it to be flat in the range $\approx {\rm U}(-\chi_\text{\tiny max}^{2g+1g},\chi_\text{\tiny max}^{2g+1g})$ with $\chi_\text{\tiny max}^{2g+1g} = 1/2$, i.e.
\be\label{2g+1gder_mod}
p_\text{\tiny pop}^{2g+1g\, {\rm mod}}(\chieff) \simeq \f{1}{2 \chi_\text{\tiny max}^{2g+1g}} 
\quad \text{for} \quad
|\chieff|\le \chi_\text{\tiny max}^{2g+1g}\,,
\ee
We do not model the drop-off observed in the tails of the $\chieff$ distribution, as it depends on the mass ratio distribution within a given model and would require additional parameters. Additionally, we notice that $2g+2g$ mergers are only affecting small tails of the BBH distribution, falling below the $y$-axis range shown in Fig.~\ref{fig:ABHdyn}, and still out of reach given the current size of the catalog.

\noindent 
{\bf \em Mass-dependent multi-generation fraction.}
We parametrise the differential contribution from $1g$ and $1g+2g$ events in the dynamical model as 
\begin{align}\label{dynmodel}
    p_\text{\tiny pop}^\text{\tiny ABH}(\chieff) 
    &
    = (1-f_g) 
    p_\text{\tiny pop}^{1g}(\chieff)
    +
    f_{g}\, 
    p_\text{\tiny pop}^{2g+1g\, {\rm mod}}(\chieff),
\end{align}
where we allow for a possible correlation with the mass parameters by assuming
\begin{align}\label{fraction_param}
    f_{g}  &= 
    f_0 + \alpha_M (M /60 M_\odot)
    - \alpha_q (q-0.5),
\end{align}
subject to the constraint imposed by $f_{g} \in [0,1]$.
We neglect the $2g+2g$ term in Eq.~\eqref{dynmodel}
since, as shown in Fig.~\ref{fig:ABHdyn}, it gives a subdominant contribution. 
This functional form is motivated by the expectation of multi-generation mergers to be populating the more massive portion of the catalog while also being associated with a smaller mass ratio.
We test the goodness of this ansatz by comparing 
Eq.~\eqref{dynmodel} with the astrophysical models from Ref.~\cite{Zevin:2020gbd}.
In particular, the GC model (and similarly for the NSC model) with $\chi_\text{\tiny b } = 0$
is very well recovered by choosing the best-fit parameters 
\begin{equation}\label{fitparamsGC}
\chimax \simeq 0.074,  
\,\,\,
f_0 \simeq 0.41,
\,\,\,
\alpha_M \simeq 0.40,
\,\,\,
\alpha_q \simeq 2.72.
\end{equation}

\begin{figure}[t!]
\centering
\includegraphics[width=0.49\textwidth]{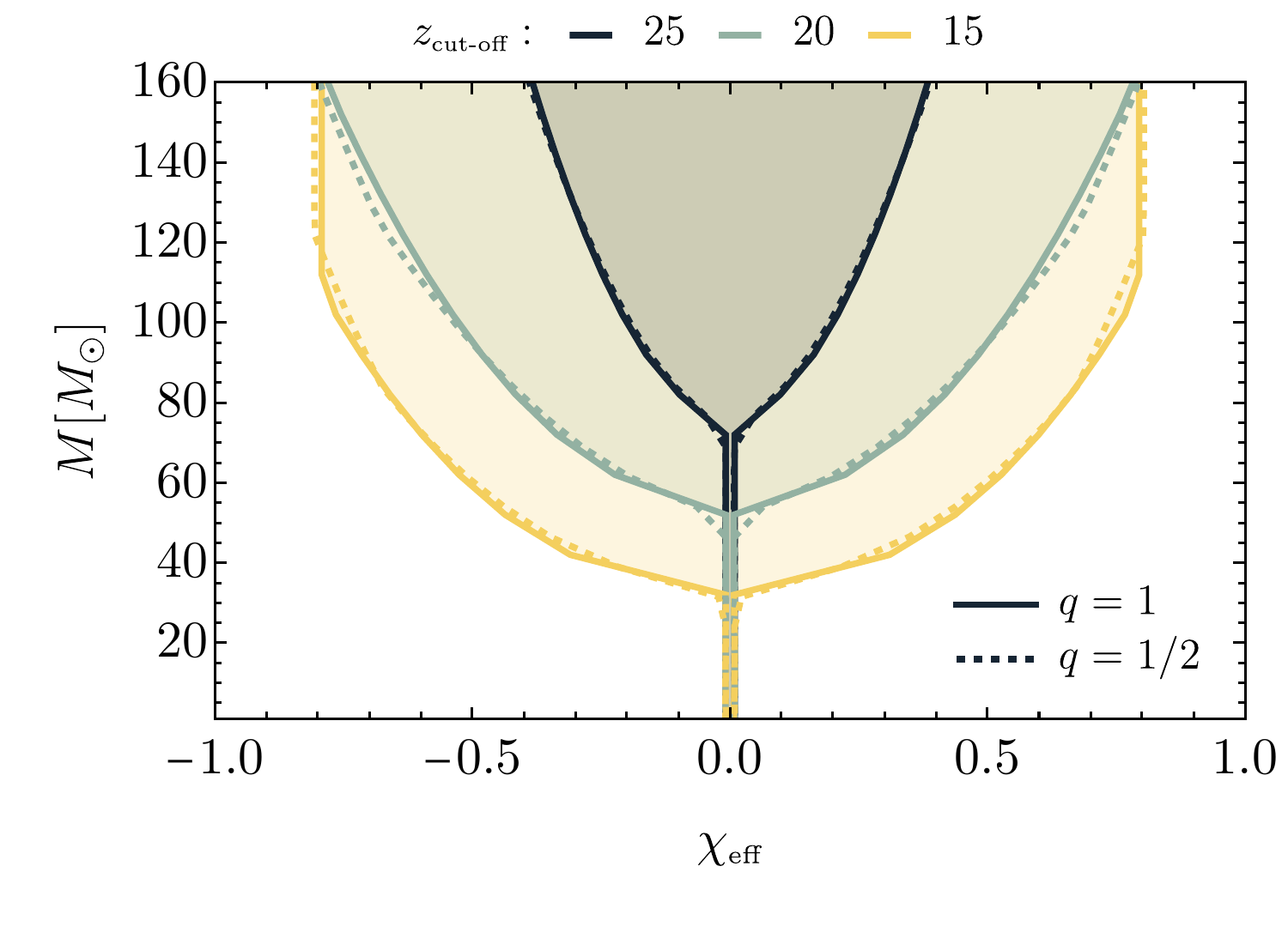}
\caption{
Distribution of $\chi_\text{\tiny\rm eff}$ (2-$\sigma$ contours) for PBH binaries as a function of the total mass $M$ and various values of $z_\text{\tiny \rm cut-off}$. In order to highlight the negligible dependence on the mass ratio, we plot the result for  $q=1$ ($q=1/2$) with solid (dashed) lines. 
}\label{fig: PBH mass_spin}
\end{figure}

\subsection{PBH model}\label{sec:th_mspins}
In this section, we review the PBH model and its predictions for the $q-\chieff$ correlation. We redirect the reader to Ref.~\cite{Franciolini:2021xbq} and references therein for more details. 
In the ``standard'' formation scenario, assumed throughout this work, PBH forms out of large Gaussian density fluctuations in the radiation dominated Universe~\cite{Sasaki:2018dmp}. This formation mechanism leads to a small ($\lesssim10^{-2}$) initial dimensionless Kerr parameter $\chi \equiv J/m^2$ (where $J$ and $m$ are the BH angular momentum and mass) at formation~\cite{DeLuca:2019buf,Mirbabayi:2019uph}. 
However, larger spins can be acquired by PBHs forming binaries through an efficient phase of accretion~\cite{DeLuca:2020qqa,DeLuca:2020fpg,DeLuca:2020bjf} prior to the reionization epoch, which was shown to affect only PBH binaries with total masses above $M\gtrsim  {\cal O}(20) M_\odot$.
While individual accretion of each PBH composing the binary is modulated by the orbital evolution~\cite{DeLuca:2020qqa},
the overall binary accretion rate is driven by the total mass of the binary determining the gravitational potential well seen by the accreting gas at the Bondi radii. As a consequence, the binary $\chieff$ inherits only a dependence on $M$ while being almost independent on the mass ratio $q$.
Finally, like in the astrophysical dynamical scenarios, spin directions of PBHs in binaries are uncorrelated and randomly distributed on the sphere~\cite{DeLuca:2020qqa} and the distribution of $\chieff$ is symmetric around zero.

Overall, PBH accretion is still affected by large uncertainties coming from the impact of feedback effects~\cite{Ricotti:2007jk,Ali-Haimoud:2017rtz}, structure formation~\cite{Hasinger:2020ptw,Hutsi:2019hlw}, and early X-ray pre-heating (e.g.~\citep{Oh:2003pm}).
In particular,  both the onset of structure formation and the reionization of the interstellar medium are expected to drastically reduce the accretion efficiency and effectively stop the impact of mass accretion on the PBH population.
Following Ref.~\cite{DeLuca:2020bjf}, to account for accretion model uncertainties, we introduce a further  hyperparameter, namely the cut-off redshift $z_\co \in [10,30]$, which gives the redshift below which accretion is assumed to be negligible.
For each value of $z_\co$ there is a one-to-one correspondence between the initial and final masses, which can be computed according to the accretion model described in detail in Refs.~\citep{Ricotti:2007jk,Ricotti:2007au,DeLuca:2020qqa,DeLuca:2020bjf}. We highlight, for clarity, that a lower cut-off is associated to stronger accretion and vice-versa. Values above $z_\co \simeq 30$ effectively correspond to negligible accretion in the mass range of interest for LVKC observations.\footnote{Notice that previous analyses comparing PBH spins with GWTC-1 data in Refs.~\cite{Fernandez:2019kyb,Garcia-Bellido:2020pwq} did not model the effect of accretion onto PBHs in binaries nor the impact of spin-dependent selection bias in GW experiments (see e.g. Ref.~\cite{Ng:2018neg}). Here we account for both effects, see Appendix~\ref{appendix_2} for details.}

To summarise, the key predictions of the primordial scenario for the distribution of $\chieff$ are the following \cite{Franciolini:2021xbq} (see also Fig.~\ref{fig: PBH mass_spin}): 
{\it i)} binaries possess negligible $\chieff$ in the ``light'' portion of the observable mass range of current ground-based detectors; 
{\it ii)}
At larger masses one expects a correlation between large binary total masses and wide distributions of $\chieff$ induced by accretion effects;
{\it iii)} The distribution of $\chieff$ is only dependent on total mass of the binary, with a negligible dependence on mass ratio (see also Fig.~\ref{fig: PBH mass_spin}). 
Therefore, for fixed primary mass $m_1$, 
the $\chieff$ distribution widens for larger $q$.
This trend with $q$ is {\it opposite} to the one observed in
dynamically formed ABH binaries. 
{\it iv)} The uncertainties in the accretion efficiency are controlled by the population hyperparameter $z_\text{\tiny cut-off}$.

\begin{table*}
\caption{Population hyperparameters ${\bm \lambda}$ 
for each model considered in this work, along with their prior distributions. We refer to a uniform distribution between two values $\theta_\text{\tiny min }$ and $\theta_\text{\tiny max}$ as ${\rm U} [\theta_\text{\tiny min },\theta_\text{\tiny max}]$.} 
\begin{tabularx}{2.05 \columnwidth}{|X||c|c||c|c|c|c||c|c|c|c||c|}
\hline
  \hline
  Model &
  \multicolumn{2}{c||}{G}   &
  \multicolumn{4}{c||}{G$_\text{\tiny corr}$}   &
  \multicolumn{4}{c||}{ABH} &
  \multicolumn{1}{c|}{PBH} 
  \\
  \hline
  Parameter & 
  $\mu$& $\sigma$ &
  $\mu_0$& $\log_{10}\sigma_0$ &$\alpha$ &$\beta$ &
  $\chi_\text{\tiny max}$ &
  $f_0$ & $\alpha_M$ & $\alpha_q$ &
  $z_\text{\tiny cut-off}$ \\
  \hline
  Prior & 
  \sppdd ${\rm U}[-1,1]$ \sppdd&\sppdd
  ${\rm U}[0,2]$ \sppdd&\sppdd
  ${\rm U}[-1,1]$ \sppdd&\sppdd
  ${\rm U}[-1.5,0.5]$ \sppdd&\sppdd
  ${\rm U}[-2.5,1]$ \sppdd&\sppdd
  ${\rm U}[-2,1.5]$ \sppdd&\sppdd
  ${\rm U}[0,1]$ \sppdd&\sppdd
  ${\rm U}[0,1]$ \sppdd& \sppdd
  ${\rm U}[0,1]$ \sppdd & \sppdd
  ${\rm U}[0,10]$ \sppdd& \sppdd
  ${\rm U}[10,30]$ \sppdd
  \\
 \hline
  \hline
\end{tabularx}
\label{tab:priors}
\end{table*}

\begin{table*}[t!]
\renewcommand{\arraystretch}{1.3}
\caption{
Mixing fractions and Bayesian evidence ratios for the various models analysed in this work. 
}
\begin{tabularx}{2.05 \columnwidth}{|X||c|c|c||c|c|c|c|}
\hline
\hline
 Model ${\cal M}$ &
 \spp   {G+ABH} \spp & 
 \spp {G+PBH} \spp   & 
 \spp {G+ABH+PBH} \spp &
 \spp   {G$_\text{\tiny corr}$} \spp & 
  \spp   {G$_\text{\tiny corr}$+ABH} \spp & 
 \spp {G$_\text{\tiny corr}$+PBH} \spp   & 
 \spp {G$_\text{\tiny corr}$+ABH+PBH} \spp
\\
\hline
Fraction $r_{\cal M}$ &
\spp $0.68^{+0.28}_{-0.41}$ \spp &
\spp $0.51_{-0.29}^{+0.25}$ \spp &
\spp ($0.37_{-0.30}^{+0.29}$,
$0.30_{-0.23}^{+0.28}$) \spp
&
-
&
\spp $0.77^{+0.20}_{-0.36}$ \spp &
\spp $0.68^{+0.20}_{-0.31}$  \spp  &
\spp ($0.34_{-0.29}^{+0.36}$,
$0.32_{-0.24}^{+0.31}$) \spp
\\
\hline
$\log_{10} {\cal B}^{\cal  M}_\text{\tiny G} \spp $ &
\spp  0.94 \spp &
\spp 0.88 \spp &
\spp 1.33 \spp
&
\spp $1.06$ \spp
&
\spp 2.15 \spp &
\spp 1.72 \spp &
\spp $2.40$ \spp
\\
 \hline
  \hline
\end{tabularx}
\label{tabbayes}
\end{table*}

To describe the distribution $p_\text{\tiny pop}^\text{\tiny PBH}(\chieff | z_\text{\tiny cut-off})$, we will adopt the analytical fit of the relation between the masses and spins predicted at low redshift (that is, $z\lesssim z_\text{\tiny cut-off}$) as a function of $z_\co \subset[10,30]$ derived in Ref.~\cite{Franciolini:2021xbq} and based on the accretion model studied in Refs.~\cite{DeLuca:2020bjf,DeLuca:2020qqa}.

\section{Searching for mass-spin correlations in GWTC-3 data} \label{sec:GWTC-3}

In this section, we report the result of the hierarchical Bayesian inference on the GWTC-3 catalog based on the spin models previously described, with the aim of searching for some signatures of the mass-spin correlation in the currently available GWTC-3 data. 
We fully report on the adopted technique in Appendix~\ref{appendix_2}, where we describe in details how we account for the LVKC detection bias using the results of the injection campaign released by the LVKC. 
In particular, it is important to take into account that larger $\chieff$ leads to larger detection probabilities~\cite{Ng:2018neg}.
We assume the mass distribution to be described by the fiducial \textsc{Power Law + Peak} model \cite{Talbot:2018cva}, whose parameters are fixed to the best fit values obtained in the recent LVKC analysis~\cite{2021arXiv211103634T}.
We report the $\chieff$ population hyperparameters for each model, along with their priors, in Table~\ref{tab:priors}.
The relevant corner plots of the posterior distributions are presented in Appendix~\ref{appendix_1} and discussed below.

\begin{figure}[t!]
\centering
\includegraphics[width=0.49\textwidth]{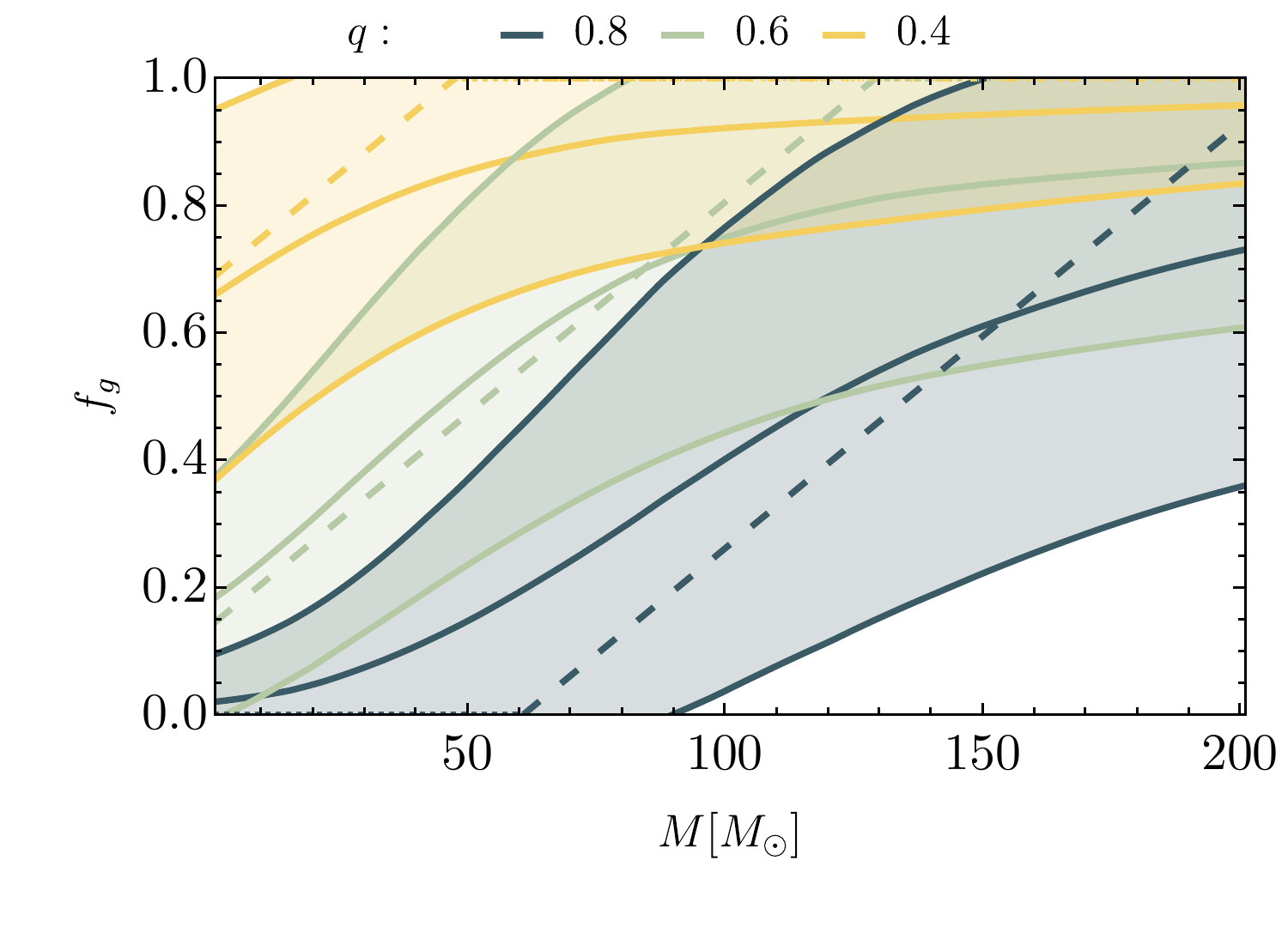}
\caption{
1$\sigma$ posterior distribution for $f_{g}$ as a function of total mass $M$ and mass ratio $q$ obtained in the G+ABH model. 
Dashed lines indicate $f_g(M)$ found with the fit in Eq.~\eqref{fitparamsGC} of the GC ($\chi_\text{\tiny b}=0$) model.
}\label{fig: pos ABH f}
\end{figure}

\subsection{Mixed population inference}
First, we separately mix either the ABH or the PBH model 
with the \textsc{Gaussian} model, 
with (G$_\text{\tiny corr}$) and without (G) the $q-\chieff$ correlation, as
\begin{align}
    p_\text{\tiny pop}(\chi_\text{\tiny eff})
    = (1- r_{\cal M}) p_\text{\tiny pop}^{\text{\tiny G}} + r_{\cal M} p_\text{\tiny pop}^{\cal M}, 
\end{align}
with ${\cal M} = {\rm ABH}$ or ${\rm PBH}$. 
In both cases, we find that the dynamical channels can account for more than half of the intrinsic binary population. We show the mixing fractions in Table~\ref{tabbayes}.

\begin{figure*}[t]
\centering
\begin{subfigure}[b]{1\textwidth}
          \centering
\includegraphics[width=0.49\textwidth]{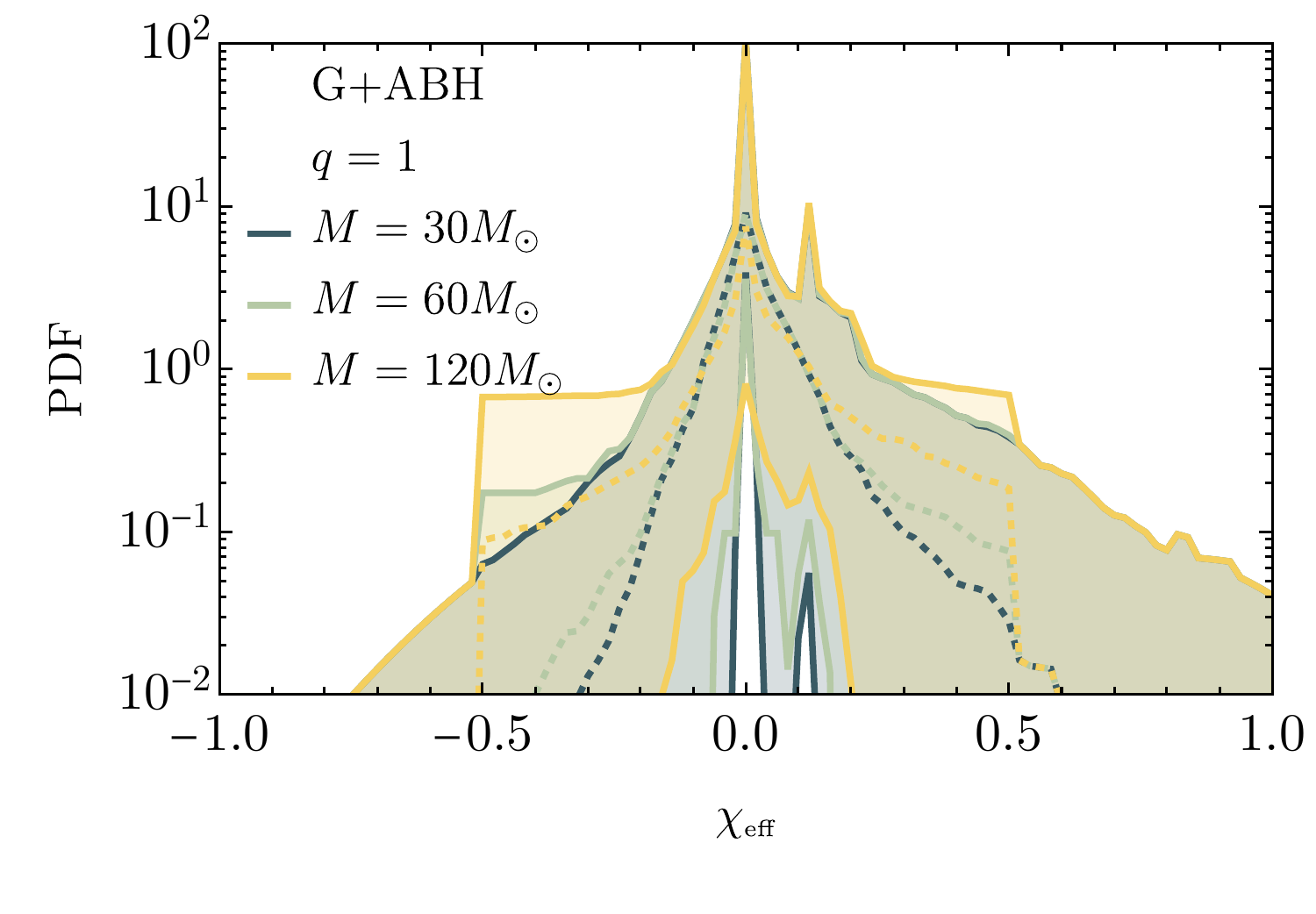}
\includegraphics[width=0.49\textwidth]{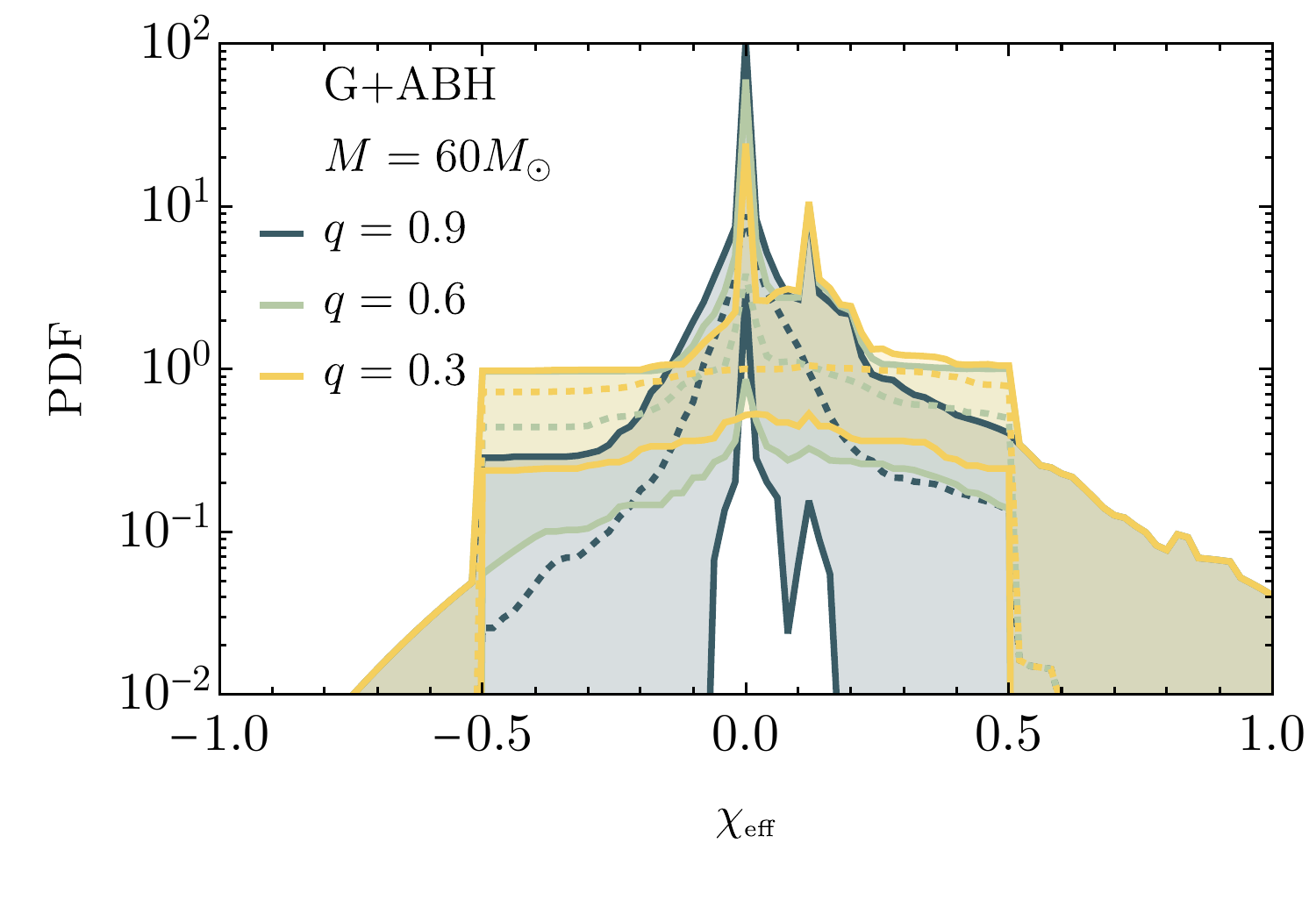}
\end{subfigure}
\begin{subfigure}[b]{1\textwidth}
          \centering
\includegraphics[width=0.49\textwidth]{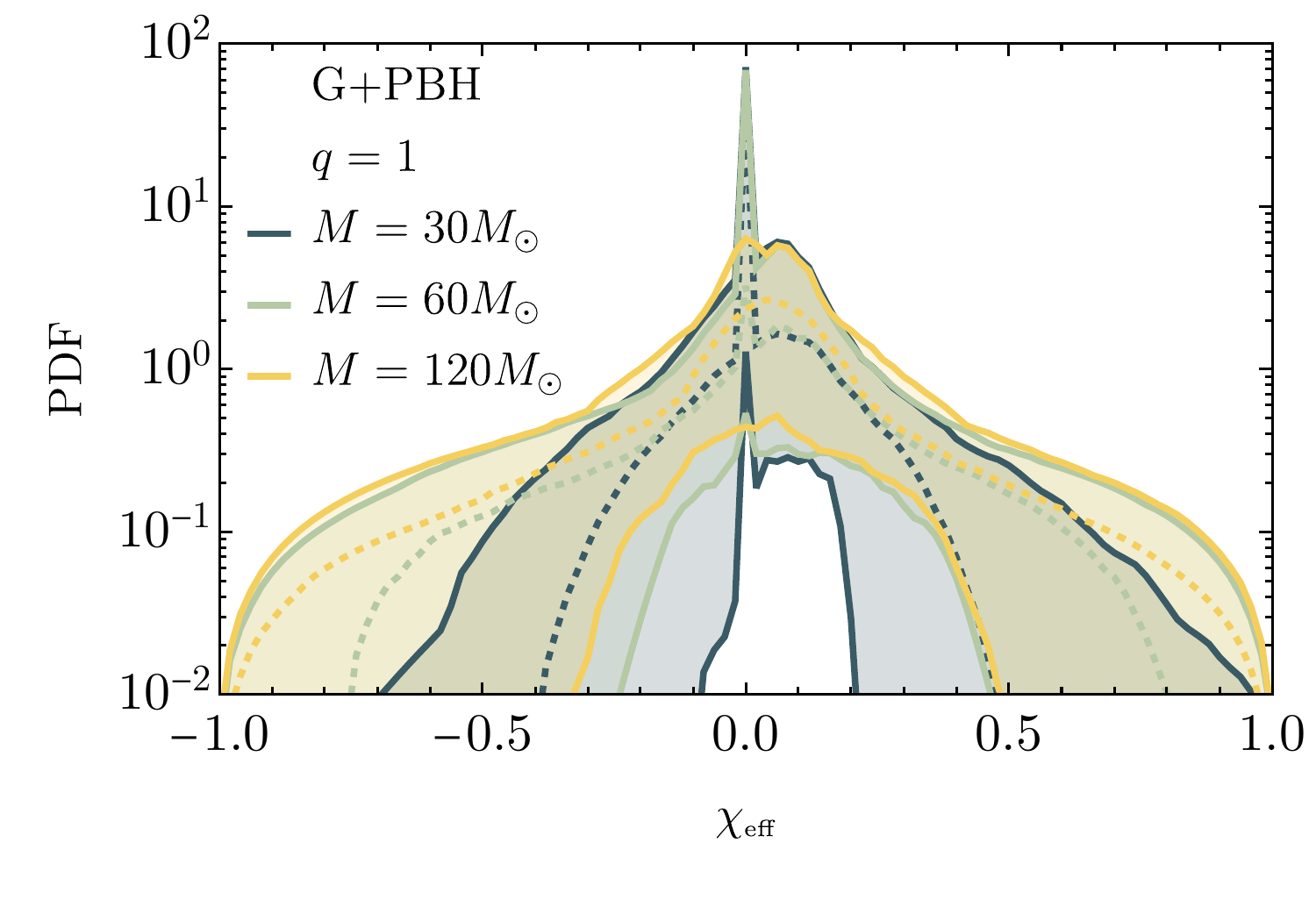}
\includegraphics[width=0.49\textwidth]{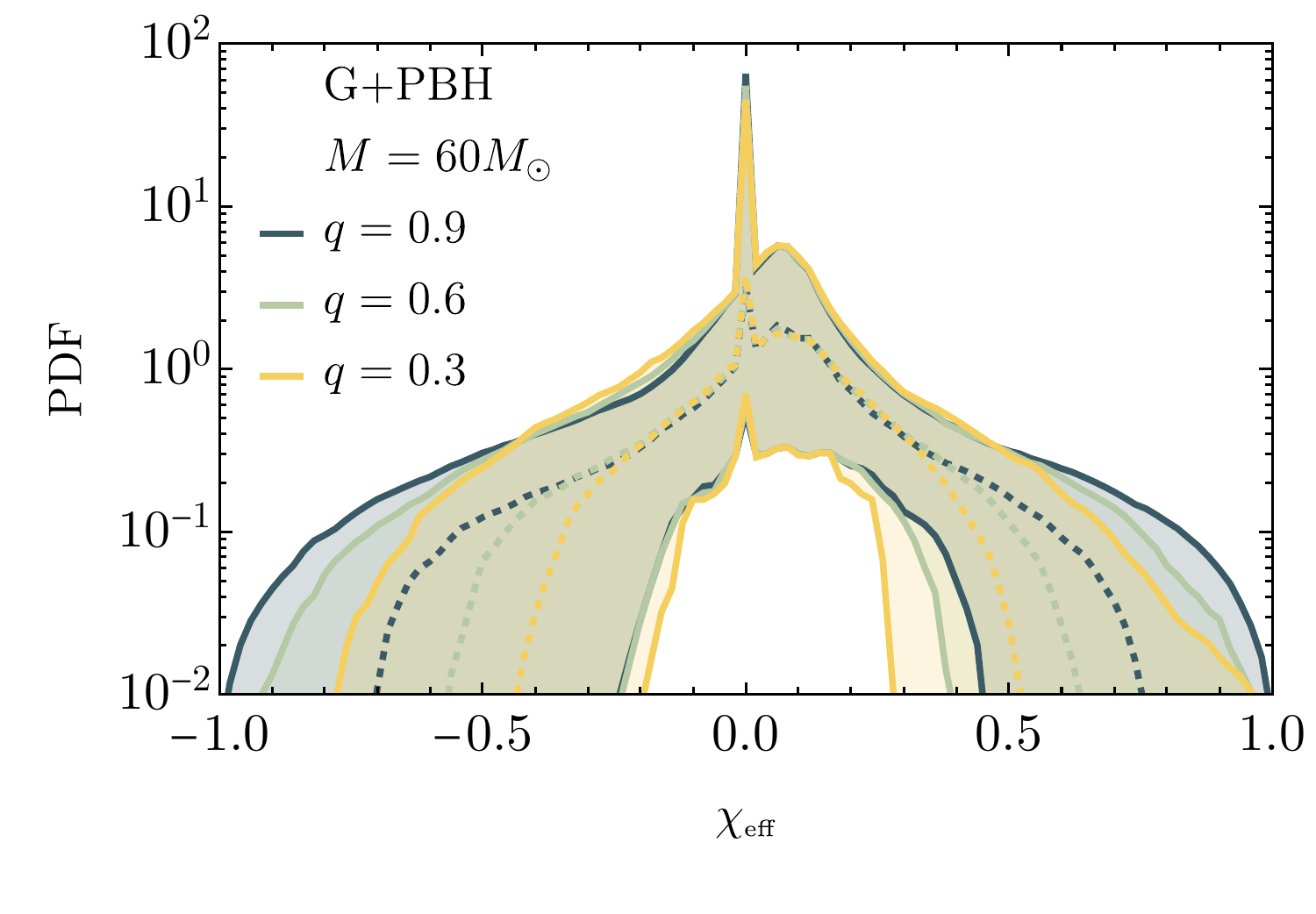}
\end{subfigure}
\caption{ 
Intrinsic distribution of $\chieff$ for various choices of total mass 
$M$ (left) and mass ratio $q$ (right).
We show results of the mixed population inference with the Gaussian model without correlations with the mass ratio. 
The colored bands indicate the $90\%$ C.L. and the dashed line the median distribution.
{\bf Top:} G+ABH scenario.
{\bf Bottom:} G+PBH scenario.
} 
\label{fig:chieffdistpl}
\end{figure*}

\begin{figure*}[t]
\centering
\begin{subfigure}[b]{1\textwidth}
          \centering
\includegraphics[width=0.49\textwidth]{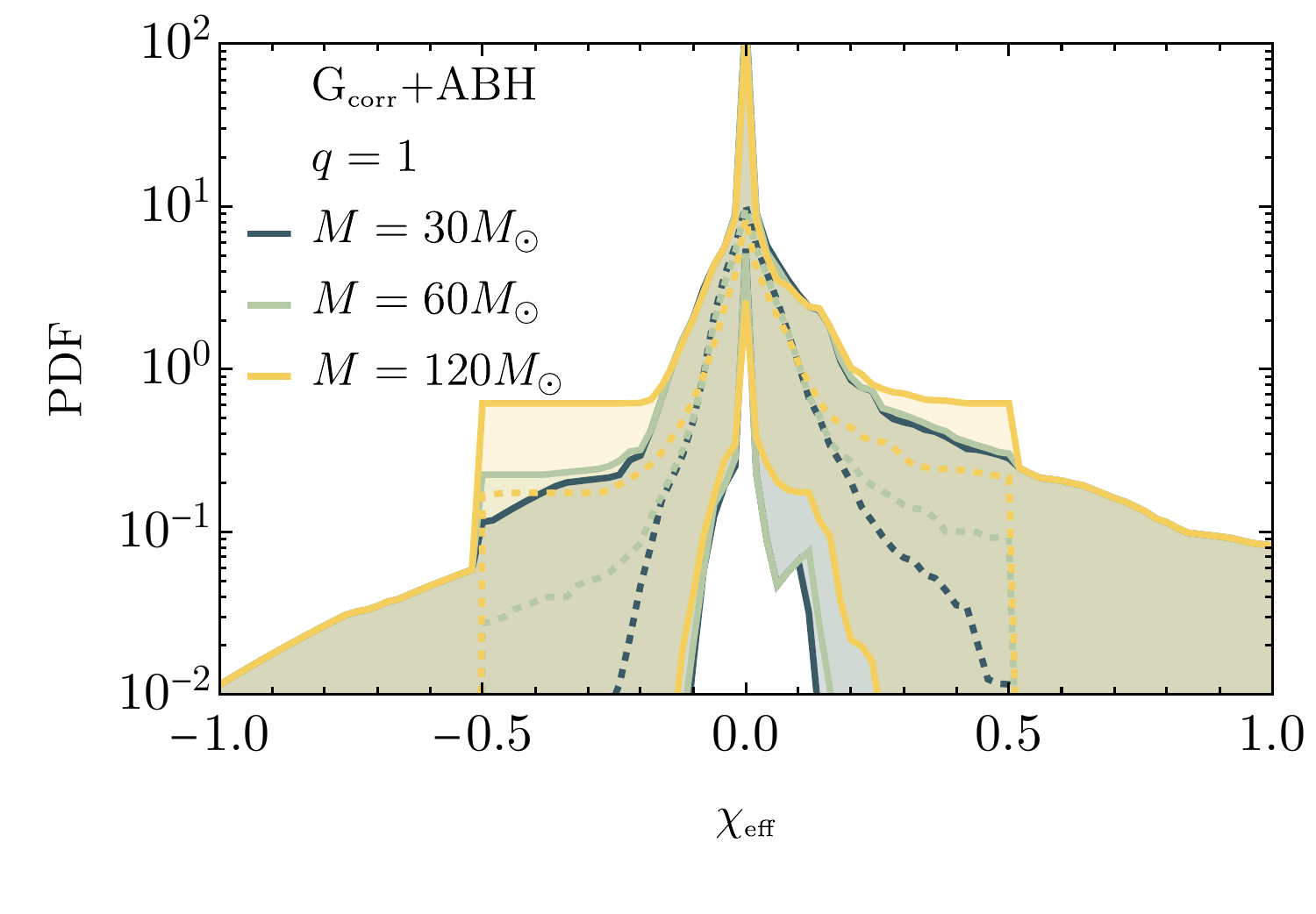}
\includegraphics[width=0.49\textwidth]{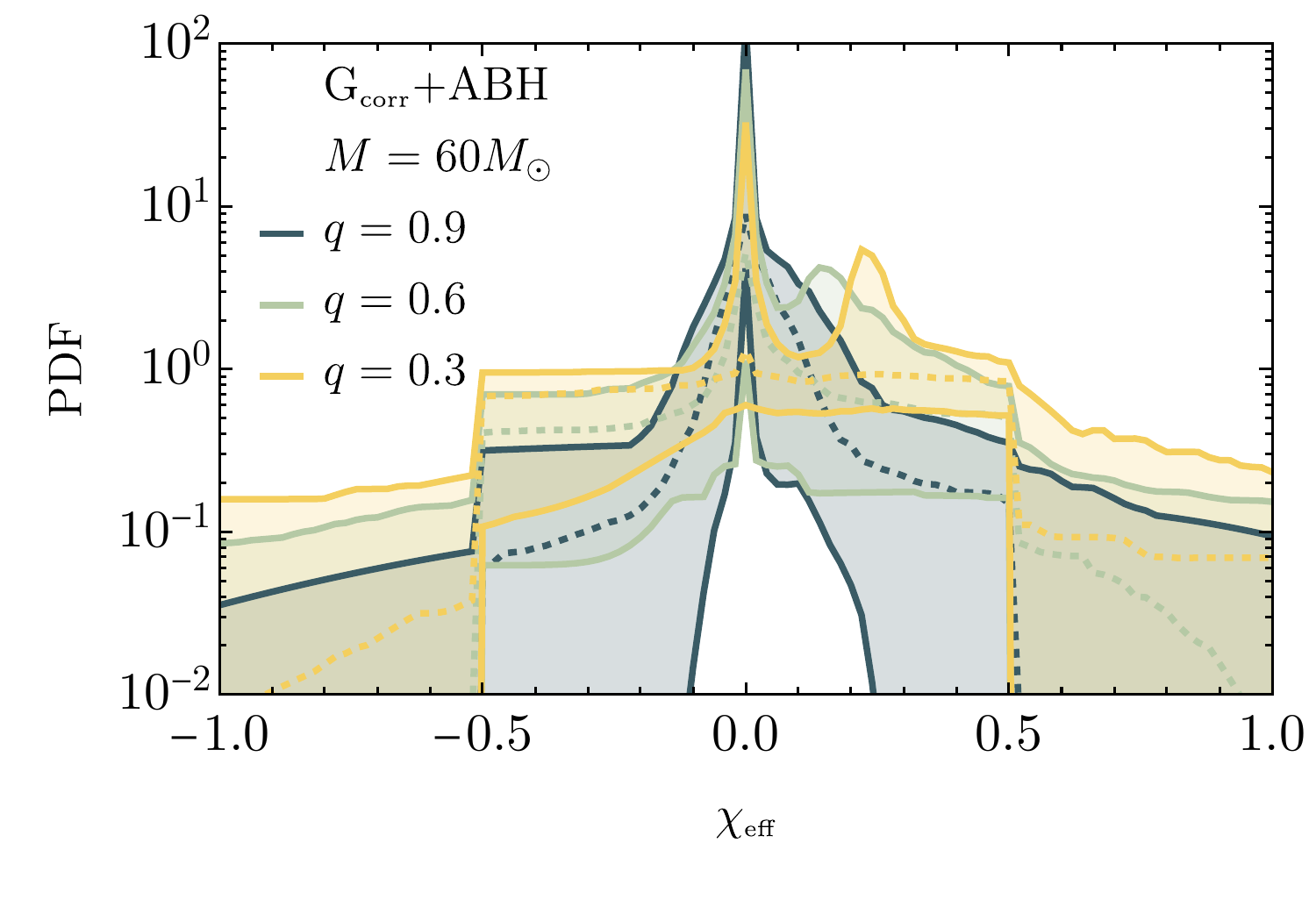}
\end{subfigure}
\begin{subfigure}[b]{1\textwidth}
          \centering
\includegraphics[width=0.49\textwidth]{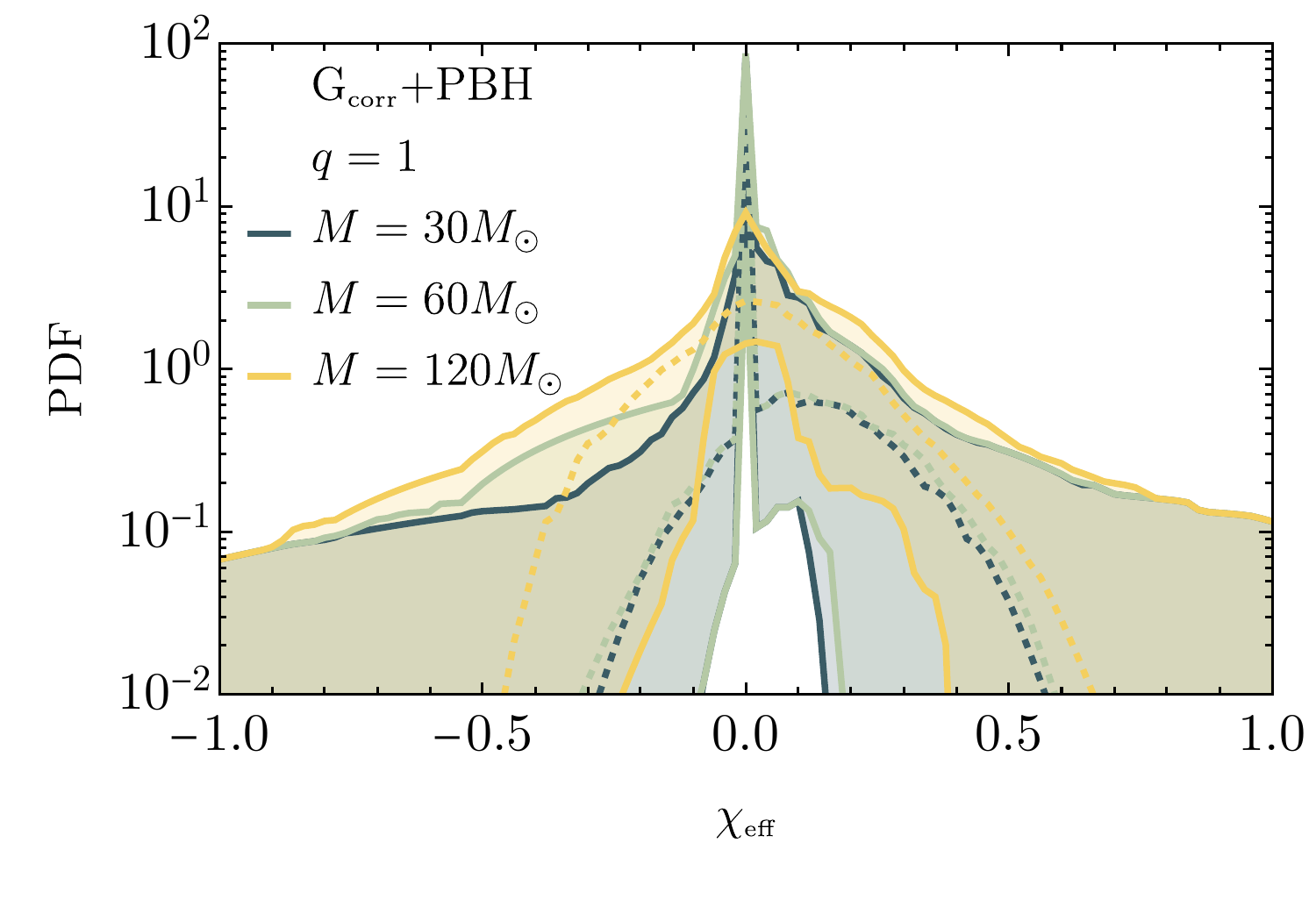}
\includegraphics[width=0.49\textwidth]{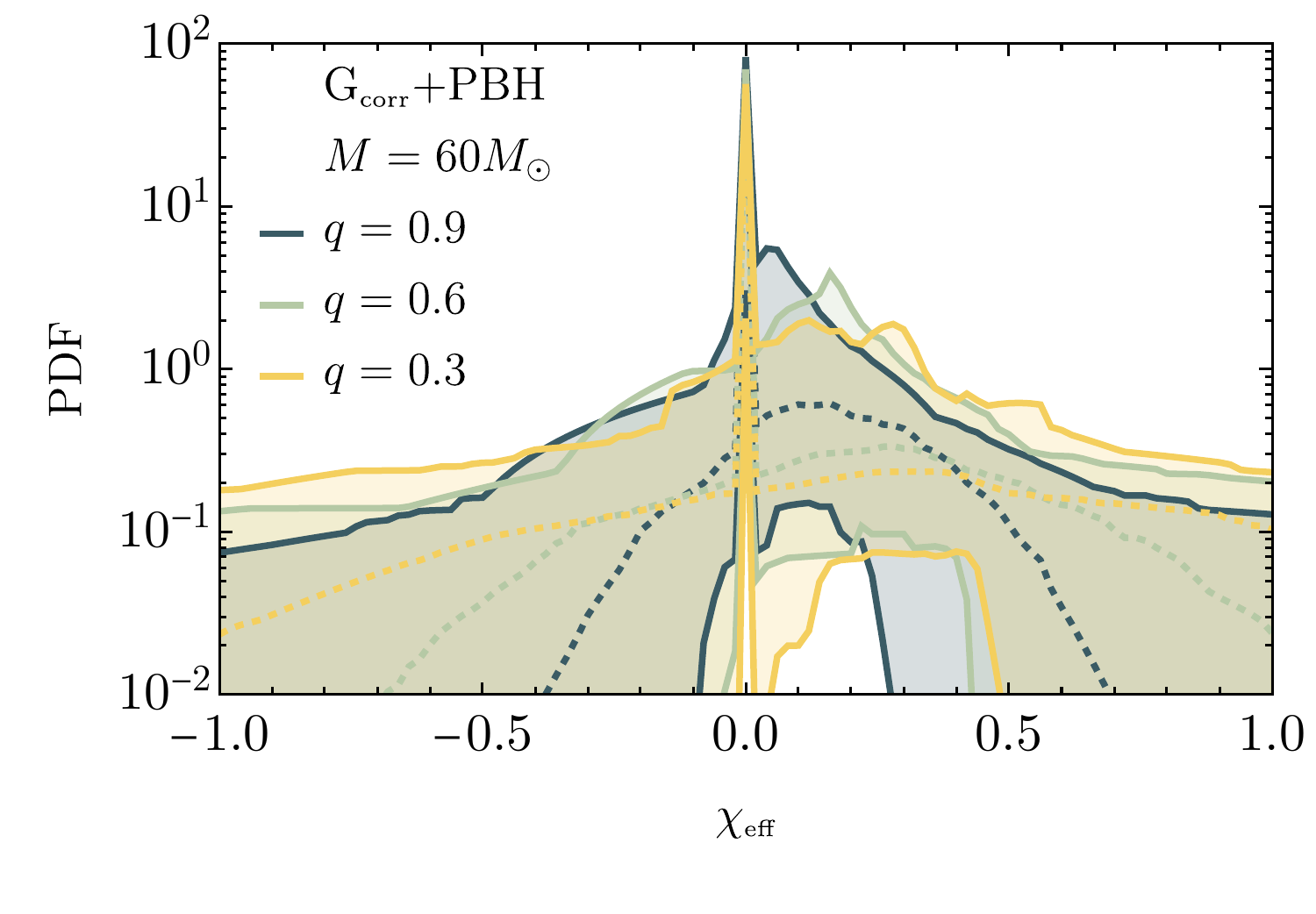}
\end{subfigure}
\caption{ 
Same as Fig.~\ref{fig:chieffdistpl} but for Gaussian models allowing for a correlation between $\chieff$ and mass ratio.
{\bf Top:} G$_\text{\tiny corr}$+ABH scenario.
{\bf Bottom:} G$_\text{\tiny corr}$+PBH scenario.
} 
\label{fig:chieffdistpl_corr}
\end{figure*}

In Fig.~\ref{fig: pos ABH f} we show
the fraction $f_{g}(M,q|f_0, \alpha_M, \alpha_q)$ of $2g+1g$ events averaged over the posterior distribution as a function of the total mass $M$ and for the G+ABH model. As one can see, although the error bars are large, the analysis naturally selects 
a fraction of second generation mergers which is growing with $M$ and with a smaller $q$. This
is very similar to what is observed in both the GC and NSC models from Ref.~\cite{Zevin:2020gbd}. Indeed, for a visual comparison, in Fig.~\ref{fig: pos ABH f} we also show the fit of those models performed with the parametrisation in Eq.~\eqref{fraction_param} (dashed curves).  

In the PBH case the fraction of binaries belonging to the 
primordial population is slightly smaller, but still compatible with a significant fraction of BBHs in the intrinsic population. 
It is noteworthy that this occurs even if the PBH model contains a single hyperparameter (as opposite to the ABH model with 4 hyperparameters, see Table~\ref{tab:priors}).
It is also interesting to notice that larger fractions are obtained with larger values of $z_\co$, which means inefficient accretion and a milder spin growth as a function of total mass $M$. On the other hand, a second peak populating the portion of the posterior with smaller $r_\text{\tiny PBH}$ is compatible with strong accretion (see more details in Appendix~\ref{appendix_1}). 
Also, the posterior distribution of 
$r_\PBH$ is incompatible with unity (see Fig.~\ref{fig:PBH+G} in Appendix~\ref{appendix_1}) as some light events in the GWTC-3 catalog have non-vanishing $\chieff$ which cannot be recovered by the PBH model, even for small $z_\text{\tiny cut-off}$ (strong accretion).
This is consistent with the conclusions of a single event analysis performed in Ref.~\cite{Franciolini:2021xbq}.

In Table~\ref{tabbayes} we report the Bayes factors comparing each mixed model with the reference \textsc{Gaussian} models.
The inclusion of a dynamical sub-population of mergers is found to be substantially favoured compared to the LVKC phenomenological model, supporting the hypothesis that mass-spin correlations as the one expected in dynamical channels (either from astrophysical or primordial scenarios) may be needed to explain the data.
Interestingly, this conclusion applies regardless of the fact that in the reference phenomenological \textsc{Gaussian} model $q-\chieff$ correlations are included or not. Indeed, as shown in Table~\ref{tabbayes}, the G$_{\rm corr}$ model is mildly favoured with respect to the standard G model, confirming the results of Ref.~\cite{Callister:2021fpo}. The inclusion of $q-\chieff$ correlations in the reference model simply shifts $\log_{10}{\cal B}_G^{\cal M}$, but the evidence obtained by adding the ABH or the PBH model relative to the reference (either G or G$_{\rm corr}$) is roughly the same\footnote{We remind that, although we normalize the Bayes factors in Table~\ref{tabbayes} to the standard G model, the quantity $\log_{10}{\cal B}_G^{\cal M}$ is additive so the relative evidence between different models can be estimated by taking the difference between two entries of the last row in the table.}.
This confirms that the \textsc{Gaussian} model with a shift towards positive values of $\chieff$ for asymmetric binaries have ``orthogonal'' characteristics with respect to the correlation induced by dynamical ABH or PBH models. 

In order to further interpret the results of the inferences, 
we show the intrinsic distribution $p(\chieff)$ in the various scenarios in 
Figs.~\ref{fig:chieffdistpl} and \ref{fig:chieffdistpl_corr}.
In all cases, we observe a sharp peak centered around zero. 
This is produced either by the ABH population of first-generation mergers with a small $\chi_\text{\tiny max}$ or by PBHs in the case of large $z_\co$ (corresponding to weak accretion and small spins). This is analogous to the potential sub-population of mergers with vanishing spin found to be compatible with the data in Ref.~\cite{2021arXiv211103634T} (see also Ref.~\cite{Galaudage:2021rkt}).
A second peak at slightly larger (and positive) values of $\chieff$ is also observed. This is generated by the Gaussian sub-population, peaking around $\mu \simeq 0.06$, as obtained by the vanilla Gaussian fit performed by the LVKC analysis \cite{2021arXiv211103634T}. 

Focusing first on mixed scenarios where the Gaussian model does not possess $\chieff$-$q$ correlations (Fig.~\ref{fig:chieffdistpl}), 
we observe that in the G+ABH model, there is a significant contribution from second-generation mergers for masses larger than $\simeq 60 M_\odot$, producing a flat contribution in the range $\simeq [-0.5,0.5]$. For smaller values of mass ratio, such contribution from second-generation mergers is even more pronounced. 
The trend in the case of a primordial subpopulation is similar if one focuses on total mass, while the population does not evolve much as a function of mass ratio, as predicted by the model in Fig.~\ref{fig: PBH mass_spin}. 
Turning now on Fig.~\ref{fig:chieffdistpl_corr}, we conclude that the features observed when the Gaussian model is extended to capture $\chieff$-$q$ correlations are similar.  The noticeable differences rely on the much wider tails of the G$_\text{\tiny corr}$ model and the shift of the peak 
of the Gaussian contribution to larger positive values of $\chieff$
for smaller values of the mass ratio, again consistently with the results of 
Refs.~\cite{Callister:2021fpo,2021arXiv211103634T}.

In the analysis presented so far, we have enforced the parameters $\alpha_M$ and $\alpha_q$
of the ABH model to be positive. This choice is motivated by the nature of the ABH dynamical 
formation channel predicting positive (negative) correlation between the total mass (mass ratio) 
of the binary and the fraction $f_g$.
One can also adopt an agnostic approach and allow these parameters to take negative values. 
We repeat the Bayesian inference of the  {G$_\text{\tiny corr}$+ABH}  scenario and find that the posterior distribution of most of the parameters is generically unaffected. 
We confidently rule out negative values of $\alpha_q$, with less than $2.4\%$ of the posterior distribution having
support at $\alpha_q<0$.
However, current data show a smaller constraining power on $\alpha_M$, as the posterior distribution contains a more sizeable tail ($< 24.6 \%$ of the posterior) at $\alpha_M<0$.

\subsection{Degeneracy between the dynamical and primordial scenario}

It is rather interesting that the two similar, but physically very different, models of mass-spin correlation considered in this work (namely ABH and PBH) are analogously preferred compared to the data. 
In order to quantify to what extent the dynamical ABH and the PBH channels are producing degenerate predictions for the distribution of $\chieff$ as a function of the masses, we also investigate a scenario in which the \textsc{Gaussian} models are mixed with \emph{both} ABH and PBH scenarios, that is
\begin{align}
    p_\text{\tiny pop}(\chi_\text{\tiny eff})
    =  r_{\text{\tiny G}} 
    p_\text{\tiny pop}^{\text{\tiny G}} 
    + r_\text{\tiny ABH} p_\text{\tiny pop}^\text{\tiny ABH}
    + r_\text{\tiny PBH} p_\text{\tiny pop}^\text{\tiny PBH}, 
\end{align}
with each mixing fraction subject to the constraint $r_{\text{\tiny G}}+r_\text{\tiny ABH}+r_\text{\tiny PBH}=1$.
The mixing fractions $r_\text{\tiny ABH}$ and $r_\text{\tiny PBH}$ turn out to be negatively correlated (see Fig.~\ref{fig:all} in Appendix~\ref{appendix_1}), with the individual fractions roughly reduced by half when compared to their counterparts in the separated analyses of the previous sections. 
Also, the evidence in favour of the G+ABH+PBH model is only $\log_{10}{\cal B} \approx 0.4$ when compared to either the G+ABH or the G+PBH cases, showing that the constraining power of the GWTC-3 is limited to disentangle highly-correlated models.
We also highlight that the posterior distribution for $(r_\text{\tiny ABH},r_\PBH)$ shown in Fig.~\ref{fig:all} confidently excludes the origin (i.e. $(r_\text{\tiny ABH},r_\PBH) \neq (0,0)$), which proves current data supports the presence of additional features beyond both the G and G$_\text{\tiny corr}$ models. 

The trends observed in this analysis are confirmed even when the \textsc{Gaussian} model is extended to allow for a positive $\chieff$
to correlate with smaller mass ratio as in Eq.~\eqref{eqcorrq}~\cite{Callister:2021fpo}.
The Bayes factors for all these scenarios  are reported in Table~\ref{tabbayes}.

\section{Conclusions and Outlook}\label{sec:conclusions}

Motivated by the recent GWTC-3 release, which substantially enlarged the number of detected GW events, we investigated whether current GW data may 
display underlying features leading back to specific BBH formation pathways. 
In particular, instead of looking at mass distributions as done in previous analyses, we focused our attention on characteristic mass-spin
correlations that are expected in dynamical BBH models, 
in particular astrophysical dynamical formation channels and a primordial-origin channel.

Overall, our results suggest the existence of correlations between masses and $\chieff$ beyond a simple phenomenological \textsc{Gaussian} model
 as the one that LVKC has adopted to describe the distribution of $\chieff$~\cite{2021arXiv211103634T}, and also beyond the extended GAUSSIAN model 
with $q-\chieff$ correlations proposed in Ref.~\cite{Callister:2021fpo}.
Interestingly, this kind of residual structure in the data appears to have characteristics which are strikingly similar to what is expected in either the astrophysical dynamical formation channels and the primordial scenario.

Having a definitive answer to the question of whether
dynamical channels produced at least a fraction of the BBHs observed to date would require a larger detection statistics and reduced measurement errors. Both these improvements will become naturally accessible in the era of third-generation GW detectors~\cite{Kalogera:2021bya}.
Along this direction, it would be interesting to 
perform a forecast of the detectability of a putative fraction of dynamical ABH/PBH binaries 
by searching for mass-spin correlation with current detectors operating at design sensitivity and with future detectors.
It would be also important to study if a larger statistics is able to break the observed degeneracy between the ABH and PBH distributions based only on $\chieff$ measurements.
If this is not the case, one may need to correlate peaks in the ABH mass distributions to wider ranges of $\chieff$ (as expected for hierarchical mergers), a feature which is absent in the PBH model, the latter producing a smoother evolution of the distribution of $\chieff$ with masses.
We plan to report some results in these directions in future work~\cite{inprep_2}.


\begin{acknowledgments}
We thank V. Baibhav for discussions on the phenomenological effective spin distributions derived in Ref.~\cite{Baibhav:2020xdf}, K. K. Y.  Ng for discussions on the GWTC-3 data release and the anonymous referee for useful suggestions.
Some computations were performed at the Sapienza University of Rome on the Vera cluster of the Amaldi Research Center funded by the MIUR program “Dipartimento di Eccellenza” (CUP: B81I18001170001).
We acknowledge financial support provided under the European Union's H2020 ERC, Starting Grant agreement no.~DarkGRA--757480, and under the MIUR PRIN and FARE programmes (GW-NEXT, CUP:~B84I20000100001), and support from the Amaldi Research Center funded by the MIUR program ``Dipartimento di Eccellenza" (CUP:~B81I18001170001).
This research has made use of data or software obtained from the Gravitational Wave Open Science Center (gw-openscience.org), a service of LIGO Laboratory, the LIGO Scientific Collaboration, the Virgo Collaboration, and KAGRA. LIGO Laboratory and Advanced LIGO are funded by the United States National Science Foundation (NSF) as well as the Science and Technology Facilities Council (STFC) of the United Kingdom, the Max-Planck-Society (MPS), and the State of Niedersachsen/Germany for support of the construction of Advanced LIGO and construction and operation of the GEO600 detector. Additional support for Advanced LIGO was provided by the Australian Research Council. Virgo is funded, through the European Gravitational Observatory (EGO), by the French Centre National de Recherche Scientifique (CNRS), the Italian Istituto Nazionale di Fisica Nucleare (INFN) and the Dutch Nikhef, with contributions by institutions from Belgium, Germany, Greece, Hungary, Ireland, Japan, Monaco, Poland, Portugal, Spain. The construction and operation of KAGRA are funded by Ministry of Education, Culture, Sports, Science and Technology (MEXT), and Japan Society for the Promotion of Science (JSPS), National Research Foundation (NRF) and Ministry of Science and ICT (MSIT) in Korea, Academia Sinica (AS) and the Ministry of Science and Technology (MoST) in Taiwan.
\end{acknowledgments}

\begin{figure*}[ht]
\centering
\includegraphics[width=0.49\textwidth]{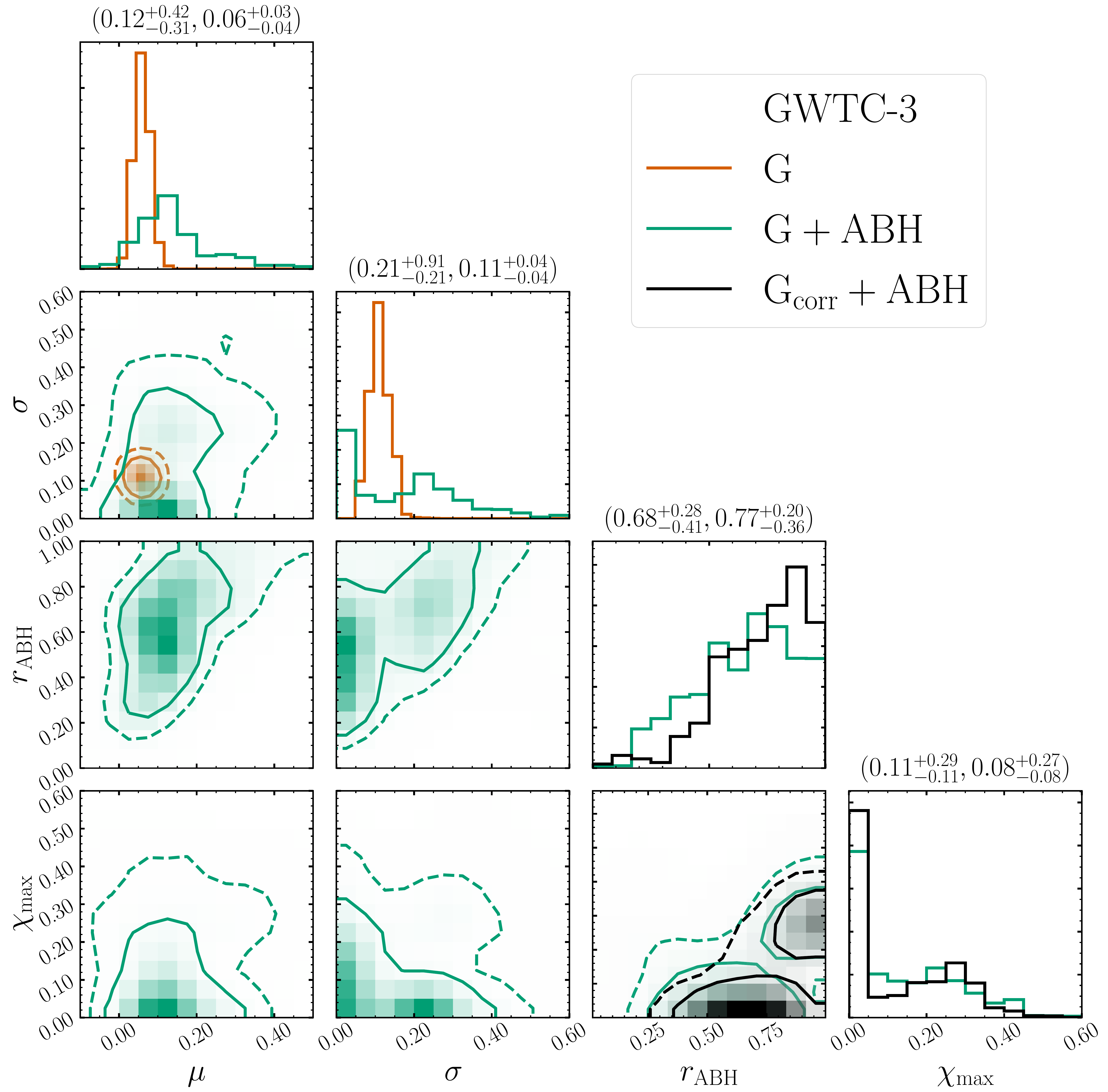}
\hspace{0.5cm}
\raisebox{.12cm}[0cm][0pt]{
\includegraphics[width=0.3675\textwidth]{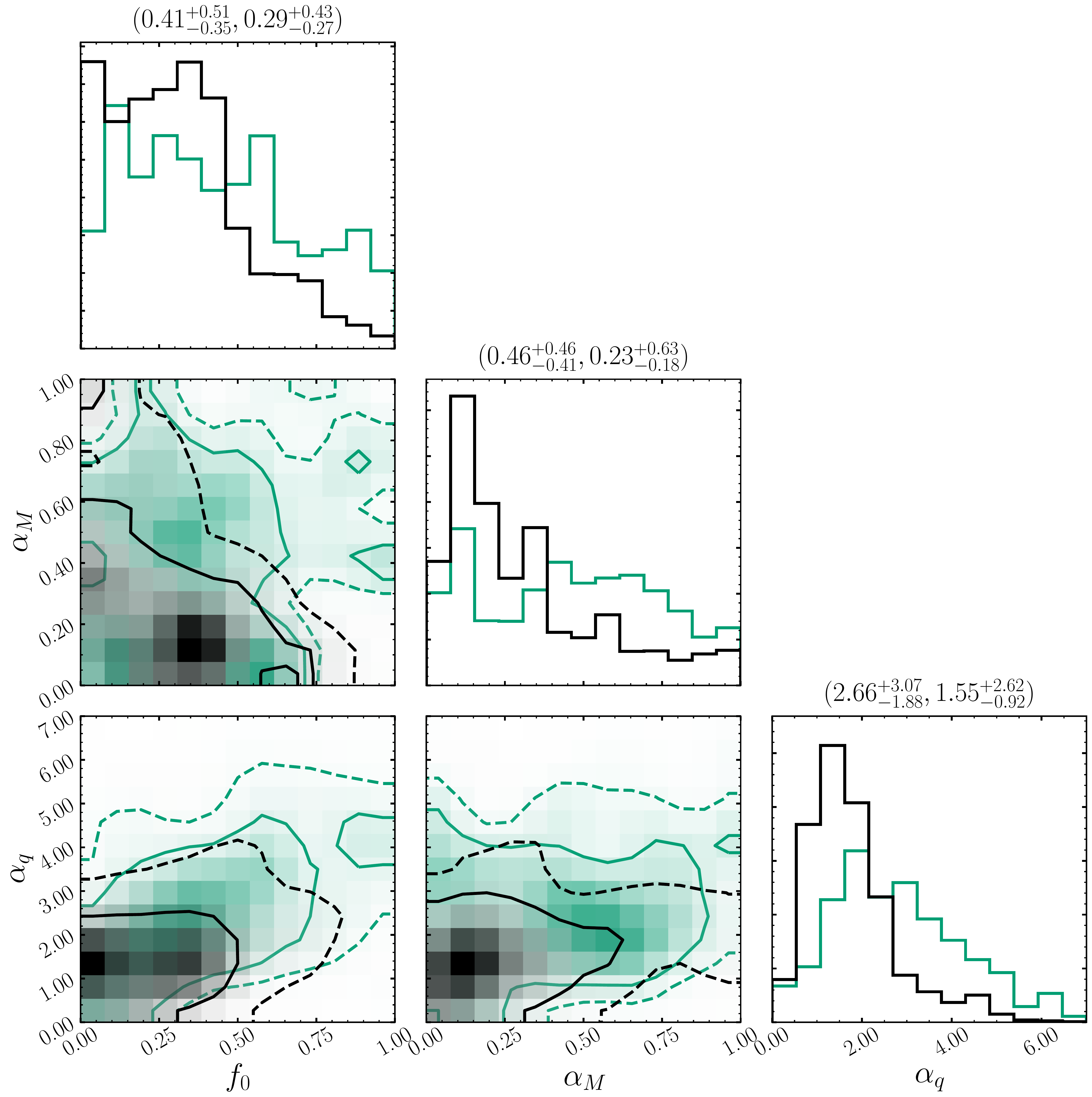}
}
\caption{ 
Posterior distribution for the mixed scenarios including an ABH sub-population. 
In orange, we show the posterior distribution for the \textsc{Gaussian} model (G) for reference. 
In green (black), we show the corresponding posterior obtained by mixing the \textsc{Gaussian} model without~\cite{2021arXiv211103634T} (with~\cite{Callister:2021fpo}) $q-\chieff$
correlations with the ABH model.} 
\label{fig:ABHdyn+G}
\end{figure*}

\begin{figure}[t]
\centering
\includegraphics[width=0.49\textwidth]{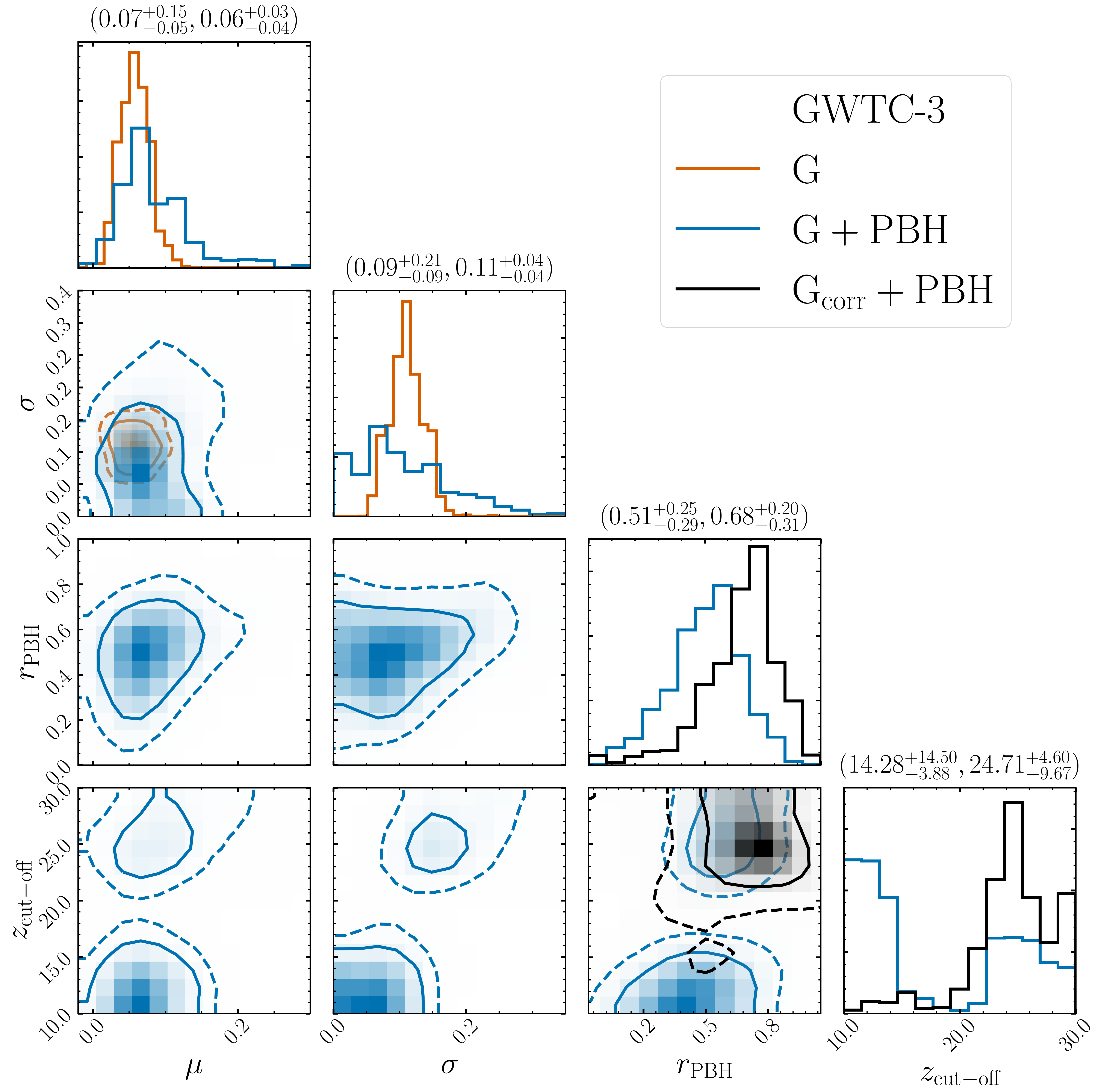}
\caption{ Same as Fig.~\ref{fig:ABHdyn+G} with a PBH sub-population. 
} 
\label{fig:PBH+G}
\end{figure}

\begin{figure*}[t]
\centering
\includegraphics[width=0.49\textwidth]{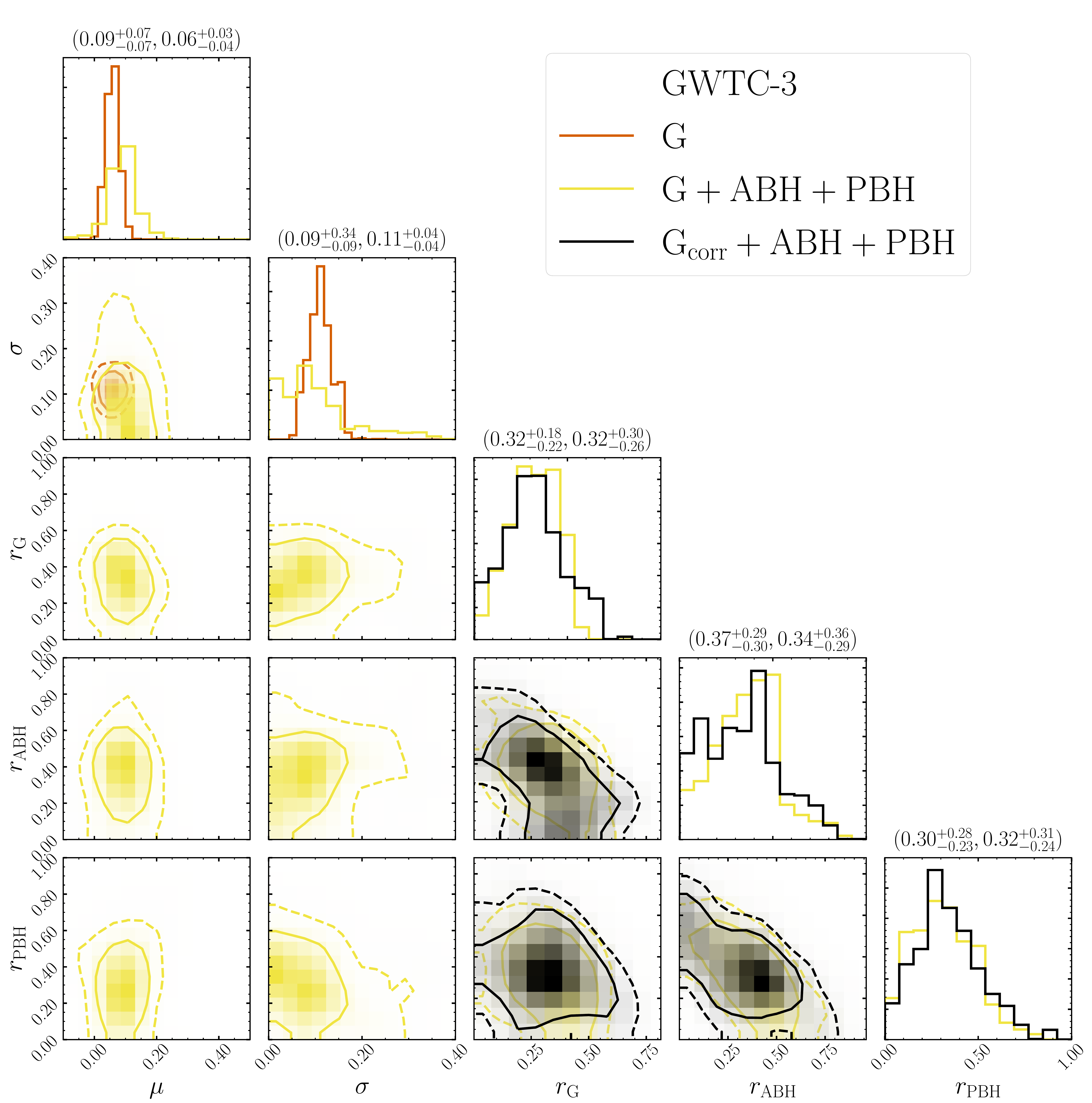}
\raisebox{-.05cm}[0cm][0pt]{
\includegraphics[width=0.48\textwidth]{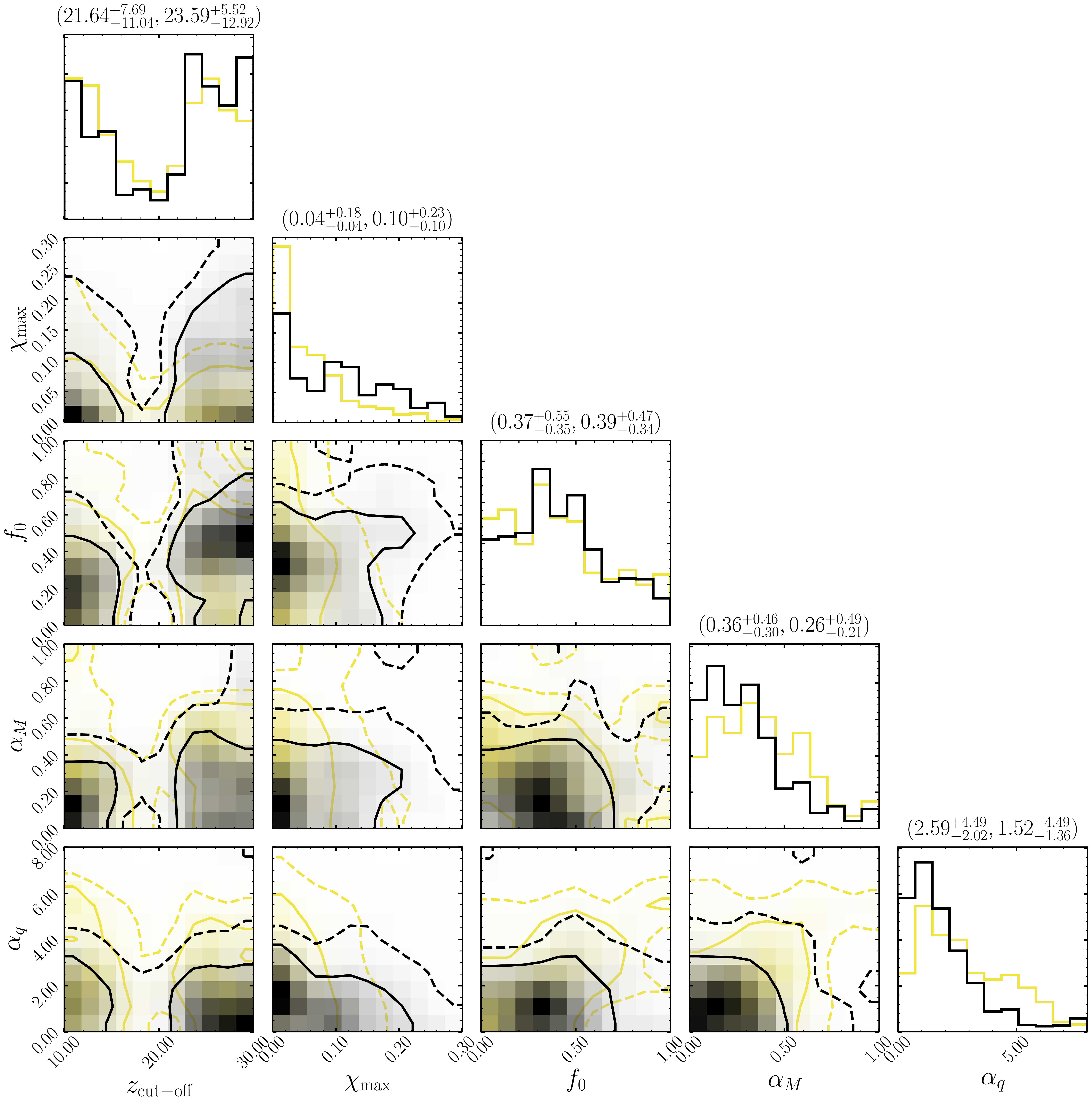}
}
\caption{ 
Same as Fig.~\ref{fig:ABHdyn+G} with both ABH and PBH sub-populations. 
} 
\label{fig:all}
\end{figure*} 

\appendix

\section{Posterior distributions for the various models}\label{appendix_1}

In this appendix we describe the features of the posterior distributions for each population inference performed in this work. 
In order to facilitate the reader, we decided to split the corner plots
with a large number of parameters in subplots showing the most relevant parameter correlations observed. 

In Fig.~\ref{fig:ABHdyn+G}, we show the posterior distributions involving the ABH channel. In red, for a visual comparison, we show the result of the inference assuming only a \textsc{Gaussian} $\chieff$ population without $q-\chieff$ correlations. 
In green, we show the result of the G+ABH mixed inference. 
As one can see, the posterior distribution for the mixing fraction $r_\text{\tiny ABH}$ is broad and peaks around $\approx 0.65$, with a tail which is compatible with unity (i.e., with all events being described by the ABH model without the need of the \textsc{Gaussian} model). Also, the distribution of $\chi_\text{\tiny max}$ (the maximum individual spin magnitude of first generation mergers) possesses a sharp peak at small values, with a second broad contribution around $\chi_\text{\tiny max}\approx 0.3$. Finally, we see that the parameters describing the fraction of second generation mergers in the ABH model are rather poorly constrained, due to the relatively small number of events populating the small mass ratio, high mass portion of the catalog.
In the same figure, we show in black the posterior distribution obtained by mixing ABHs with the G$_\text{\tiny corr}$ model. As one can appreciate, the posterior distribution is overall very similar, confirming that the correlation between small $q$ and {\it positive} values of $\chieff$ is rather orthogonal to the one expected in the ABH model, where asymmetrical binaries are correlated with larger widths of the distribution of $\chieff$. 
This is also supported by the Bayes factors reported in Table~\ref{tabbayes}, 
which show
$
\log_{10}{\cal B}_\text{\tiny G}^\text{\tiny G+ABH} 
\approx 
\log_{10}{\cal B}_{\text{\tiny G}_\text{\tiny corr}}^{\text{\tiny G}_\text{\tiny corr}+\text{\tiny ABH}} 
\approx 
1
$.
The only relevant difference is the slight shift of $\alpha_q$ to smaller values
when one allows for correlation with the mass ratio in the Gaussian model.

In Fig.~~\ref{fig:PBH+G}, we show the results involving the PBH model. 
When the \textsc{Gaussian} model is mixed with PBHs, we observe that the PBH population is able to recover a large fraction of the events, with two distinct peaks corresponding to small (large) values of $z_\text{\tiny cut-off}$. 
In particular, the portion of the posterior with larger $r_\text{\tiny PBH}$
is correlated with a larger $z_\text{\tiny cut-off}$ and milder accretion, while 
a smaller fraction of events may be compatible with stronger accretion and smaller $z_\text{\tiny cut-off}$. 
It is interesting to notice that the inclusion of $q-\chieff$ correlations in the \textsc{Gaussian} model seems to suppress the posterior 
in the small $z_\text{\tiny cut-off}$ region. This, however, does not result in a relevant difference in the Bayesian evidence and one finds
$
\log_{10}{\cal B}_\text{\tiny G}^\text{\tiny G+PBH} 
\approx 
\log_{10}{\cal B}_{\text{\tiny G}_\text{\tiny corr}}^{\text{\tiny G}_\text{\tiny corr}+\text{\tiny PBH}} 
\approx 
0.8
$.

Finally, in Fig.~\ref{fig:all}, we plot the posterior distributions for the mixed G/G$_\text{\tiny corr}$+ABH+PBH models. 
Here again, we indicate in black the result obtained assuming a \textsc{Gaussian} correlated model. 
The most noticeable feature of the posterior distribution is the apparent negative correlation between the ABH and PBH mixing fraction, as displayed in the $(r_\text{\tiny ABH},r_\text{\tiny PBH})$ plane. Additionally, the fraction of G events appears to be mostly uncorrelated with the PBH channel, as displayed in the $(r_\text{\tiny G},r_\text{\tiny PBH})$ panel of Fig.~\ref{fig:all} (left).
Overall, the posterior distribution for the hyperparameters of each sub-population are found to be similar to the one obtained in the single mixed scenarios G/G$_\text{\tiny corr}$+ABH or G/G$_\text{\tiny corr}$+PBH. 
Analogously to the result obtained in the G/G$_\text{\tiny corr}$+ABH case, the inclusion of $q-\chieff$ correlations in the \textsc{Gaussian} model reduces the mean value of $\alpha_q$.

\section{Hierarchical population inference}\label{appendix_2}

In this appendix we summarize the setup we adopt for the hierarchical population inference. For more details, we refer the reader to more comprehensive reviews of the topic~\cite{Thrane:2018qnx,Vitale:2020aaz,2019MNRAS.484.4008G,Mandel:2018mve}.

\subsection{Hyperparameter posterior}

The aim of the inference is to produce posterior distributions for the hyperparameters of a model ${\cal M}$ which is assumed to explain the GW dataset, alongside the corresponding evidence $Z_{\cal M}$ allowing for statistical model comparisons. 
The LVKC's Gravitational Wave Open Science Center~\cite{Vallisneri:2014vxa,GWOSCref} provides the output of the parameter estimation process performed for each GW signal as a collection of posterior distributions for the parameters describing the properties of individual merger events, i.e. $p({\bm \theta}|{\bm d}_i)$, where ${\bm \theta}$ indicates the binary {\it event parameters} such as masses, spins, and redshift. The index $i$ runs over all the detected GW events while, in our analysis, we consider ${\bm \theta} = (m_1,m_2,\chi_\text{\tiny eff},z)$ as our set of intrinsic binary parameters.

In order to compute the number of events produced in a given model, one needs to compute 
\begin{equation}
    N(\bm \lambda)= 
    T_\text{\tiny obs} R({\bm \lambda})
    \int 
    \d {\bm \theta} 
    p_\text{\tiny pop}(\bm{\theta}|\bm{\lambda})
    ,
\end{equation} 
where $R({\bm \lambda})$ is the intrinsic merger rate of the model ${\cal M}$,  $p_\text{\tiny pop}(\bm{\theta}|\bm{\lambda})$ is the \textit{population likelihood}, corresponding to the probability of having a binary with parameters $\bm{\theta}$ within the model ${\cal M}$ characterised by hyperparameters $\bm{\lambda}$, and $T_\text{\tiny obs}$ is the duration of the various LVKC observing runs.
We account for selection effects caused by the finite sensitivity of the detectors through the factor $0 \leq p_{\textrm{det}}(\bm \theta)\leq 1$, defining the probability that an event with parameters $\bm \theta$ would be detectable at the LVKC facilities. 
The observable number of events is thus defined as 
\begin{equation}
    N_\text{\tiny det}(\bm \lambda)= 
    T_\text{\tiny obs} R({\bm \lambda})
    \int 
    \d {\bm \theta} \,
    p_{\textrm{det}}(\bm \theta) \,
    p_\text{\tiny pop}(\bm{\theta}|\bm{\lambda})
    .
\end{equation}

Given a vector of hyperparameters $\bm{\lambda}$ (or {\it population parameters}) describing the model ${\cal M}$, 
a hierarchical Bayesian analysis produces posterior distributions inferred from the data to be
\begin{align}
p(\bm{\lambda}|\bm{d}) \propto \pi(\bm{\lambda})\int p(\bm{d}|\bm{\theta})p_\text{\tiny pop}(\bm{\theta}|\bm{\lambda})\d \bm{\theta},
\label{eq:HBA}
\end{align}
where $p(\bm{d}|\bm{\theta})$ is the single-event likelihood and $\pi(\bm{\lambda})$ is a prior on the model hyperparameters. 
In particular, the population posterior can be explicitly defined as
\begin{align}
\frac{p({\bm \lambda}|{\bm d})}{\pi({\bm \lambda}) }
\propto
e^{- N_\text{\tiny det} ({\bm \lambda})} N({\bm \lambda})^{N_\text{\tiny obs}}
\prod_{i=1}^{N_\text{\tiny obs}} 
\int {\rm d}{\bm \theta_i} 
\frac{p({\bm \theta_i}|{\bm d})  p_\text{\tiny pop}({\bm \theta_i}|{\bm \lambda})}{\pi({\bm \theta_i})} \,,
\label{eq:posterior}
\end{align}
where the prefactor introduces the standard terms describing the statistics of an inhomogeneous Poisson process (see e.g. Refs.~\citep{2004AIPC..735..195L,2018PhRvD..98h3017T,2019MNRAS.486.1086M,Thrane:2018qnx} for detailed derivations) and $\pi ({\bm \theta}_i)$ is the prior distribution over the intrinsic parameters adopted by the LVKC when performing the parameter estimation for each individual event.
As we are not interested in the overall binary merger rate, but rather in the population mixing parameters $r_{\cal M}$ as defined in the text, we can marginalise over $N(\bm \lambda)$ assuming a logarithmic prior. This yields~\cite{2018ApJ...863L..41F,2019MNRAS.486.1086M,Vitale:2020aaz}
\begin{equation}
p({\bm \lambda}|{\bm d}) \!\propto\! \pi({\bm \lambda})  \prod_{i=1}^{N_\text{\tiny obs}} \! \int \!\frac{p({\bm \theta_i}|{\bm d})}{\pi({\bm \theta_i})}
\frac{p_\text{\tiny pop}({\bm \theta_i}|{\bm \lambda})}{\alpha(\bm \lambda)}
{\rm d}{\bm \theta_i} \,,
\label{eq:posteriormarg}
\end{equation}
where $\alpha(\bm \lambda)$ is the fraction of events one would detect given a population (also known as the selection bias)
\begin{align}
\alpha(\bm \lambda) = \int p_\text{\tiny pop}({\bm \theta'}|{\bm \lambda}) p_\text{\tiny det}(\bm \theta') \d \bm \theta' = \frac{N_\text{\tiny det}(\bm \lambda)}{N(\bm \lambda)}.
\label{eq:selectionFunction}
\end{align}

In order to speed up the computations, the integral in Eq.~\eqref{eq:posterior} is typically evaluated by using importance sampling, i.e. by computing the expectation value of the prior-reweighted population likelihood by 
turning the integral into a discrete sum over the samples of the event posterior probability distribution function, which means
\begin{align}
p(\bm{\lambda}|\bm{d}) \propto 
\frac{\pi(\bm{\lambda})}{\alpha({\bm \lambda})}
\prod_{i=1}^{N_\text{\tiny obs}}\frac{1}{{\cal S}_i}\sum_{j=1}^{{\cal S}_i} \frac{p_\text{\tiny pop}(^j\bm{\theta}_i|\bm{\lambda})}{\pi(^j\bm{\theta}_i)},
\label{eq:populationPosterior_discrete}
\end{align}
where $j$ labels the $j$-th sample of the $i$-th event.
We sample Eq.~\eqref{eq:populationPosterior_discrete} using the MCMC package \texttt{emcee}~\cite{Foreman-Mackey:2012any}.

Given a model ${\cal M}$, the evidence is defined as the marginal population likelihood computed as the integral of the population posterior $p({\bm \lambda}| {\bm d})$, i.e.
\begin{equation}
Z_{\cal M} \equiv \int \d {\bm \lambda} \, p(\bm{\lambda}|\bm{d}).
\end{equation}
 We compute the evidence for each model from the posterior data following Ref.~\cite{NR1994}.
In other words, the evidence is a measure of the support for a given model given the data ${\bm d}$. One can then compare different models by computing the so-called Bayes factors
\begin{equation}
{\cal B}^{{\cal M}_1}_{{\cal M}_2} \equiv \frac{ Z_{{\cal M}_1}}{Z_{{\cal M}_2}}.
\end{equation}
According to Jeffreys' scale criterion~\cite{Jeffreys}, a Bayes factor larger than $(10,10^{1.5},10^2)$ would imply a strong, very strong, or decisive evidence in favour of model ${\cal M}_1$ with respect to model ${\cal M}_2$ given the available dataset.

\subsection{Selection bias}

In Eq.~\eqref{eq:populationPosterior_discrete}, 
the selection bias $\alpha(\Lambda)$
quantifies the fraction of events that are expected to 
overcome the detection threshold, given a population 
model ${\cal M}$ characterised by the hyperparameters  ${\bm \lambda}$.
The detection efficiency has the crucial role of correcting  for the inevitable selection effects which may systematically bias the result of the hierarchical population inference.
Following the procedure described in the LVKC publications, we estimate $\alpha({\bm \lambda})$ using the injection campaign reported in Ref.~\cite{Abbott:2020gyp},\footnote{ 
The injection campaign was extended with the inclusion of the O3b run. For simplicity, we only adopt the one released in Ref.~\cite{Abbott:2020gyp}. As we only focus on the shape of the BBH spin distribution, and in particular on $\chieff$, while we marginalise over the rate, we do not expect this to affect our result. }
selecting successfully found injections (with recovered false alarm rates below one per year in at least one pipeline) and reweighting to the proposed population with hyperparameters ${\bm \lambda}$ as 
    \begin{equation}
    \alpha({\bm \lambda}) = \frac{1}{N_\text{\tiny inj}} 
    \sum_{j=1} ^{N_\text{\tiny found}}
    \frac{p_\text{\tiny pop}({\theta_j}|{\bm \lambda})}
    {p_\text{\tiny inj}({\theta_j})}
        ,
    \label{eq:sel-effects}
    \end{equation}
where $N_\text{\tiny inj}$ is the total number of injections (including those that are not recovered) and $p_\text{\tiny inj}(\theta)$ is the reference distribution from which injections were drawn.
In particular,  the injected distribution of masses follows $p_\text{\tiny inj}(m_1) \propto m_1^{-2.35}$ for $2\,M_\odot \leq m_1 \leq 100\,M_\odot$ and $p_\text{\tiny inj}(q|m_1) \propto q^2$
while having aligned component spins
($\alpha_1 = \alpha_2= 0 $ or $\alpha_1 = \alpha_2= \pi $)
uniformly distributed in the range $\chi_{z,i} \equiv \chi_i \cos\alpha_i \subset [-1,1]$.

\subsection{Reference mass model}
In the analysis performed in the main text, we are only interested in constraining the parameters describing the distribution of the effective inspiral spin $\chi_\text{\tiny eff}$. The selection function, however, is also dependent on the mass distribution of a given model. Therefore, we decided to fix the masses in our population model to the reference model called \textsc{Power Law + Peak}~\cite{Talbot:2018cva} and adopted by the LVKC analyses~\cite{2021arXiv211103634T}. 
This model assumes that the primary BBH masses are described as a mixture of 
 a power law 
 \begin{equation}
     P(m_1|\lambda,m _\text{\tiny min},m_\text{\tiny max}) \propto m_1^\lambda
 \end{equation} 
 and a Gaussian peak
 \begin{equation}
 N(m_1|\mu_m,\sigma_m,m _\text{\tiny min},m_\text{\tiny max}) \propto \exp\left[-\frac{(m_1-\mu_m)^2}{2\sigma_m^2}\right]    
 \end{equation}
 normalized to unity across the range $ m _\text{\tiny min} \leq m_1 \leq m_\text{\tiny max}$. 
 The mixing fraction between the two components is dictated by $\lambda_\text{\tiny peak}$ as
\begin{align}
    p_\text{\tiny pop}(m_1) 
    &=
    (1-\lambda_\text{\tiny peak})
    \,P(m_1) 
    + \lambda_\text{\tiny peak} N(m_1).
    \label{eq:pm}
\end{align}
We describe the population distribution of mass ratios via a power law as
    \begin{equation}
    p(q|m_1,\gamma) \propto q^{\gamma},
    \label{eq:pq}
    \end{equation}
constrained within the range $m_\text{\tiny min}/m_1 \leq q \leq 1$.
Additionally, the reference model assumes a redshift distribution of sources that is proportional to the differential comoving volume ${\d V_c}/{dz}$ with an evolution of the merger rate at high redshift (see e.g. Ref.~\citep{2018ApJ...863L..41F}),
\begin{equation}
    p(z|\kappa) \propto \frac{1}{1+z} \frac{dV_c}{dz} \left(1+z\right)^\kappa.
    \label{eq:pz}
\end{equation}
The additional factor of $(1+z)^{-1}$ in Eq.~\eqref{eq:pz} converts a uniform-in-time source-frame distribution to our detector frame.
We fix the hyperparameters of the model, i.e. ($\lambda_\text{\tiny peak},\lambda,\gamma,\kappa,m_\text{\tiny min},m_\text{\tiny max},\mu_m,\sigma_m$), to the mean values obtained in the LVKC analysis performed in Ref.~\cite{2021arXiv211103634T}.
While evidence of an additional substructure features on top of the \textsc{Power Law + Peak} coarse grained-model was found by LVKC~\cite{2021arXiv211103634T} (see also Refs.~\cite{Tiwari:2020otp,Edelman:2021zkw,Tiwari:2021yvr,Li:2022jge}) we do not expect our results to be affected by potential systematic effects in our choice of benchmark mass model, as it was also shown to be the case in Ref.~\cite{Callister:2021fpo}.

\subsection{The GWTC-3 dataset}

Among all the binary events included in the GWTC-3 catalog, we use the same subset selected for the population analysis in Ref.~\cite{2021arXiv211103634T}. 
In particular, focusing on the population of binaries which can be confidently interpreted as BBHs, we do not include events where the secondary binary component has mass smaller than $3 M_\odot$. This implies, in particular, that we do not include GW170817, GW190425, GW190814.
Additionally, we disregard events with a false-alarm-rate larger than the threshold of $1 \,{\rm yr}^{-1}$. This leaves us with $69$ event.
\footnote{We do not include in our analysis the additional events discovered by independent searches performed by various groups outside the LVKC collaboration (e.g.~\cite{Nitz:2021zwj,Olsen:2022pin}), and leave such a task for future work. This choice allows us to directly compare our results to the population analyses performed by the LVKC in Ref.~\cite{2021arXiv211103634T}. }

We adopt the \texttt{Overall\_posterior} samples provided in~\cite{PEresult_GWTC1} for the $10$ considered events in GWTC-1, the \texttt{PrecessingSpinIMRHM} provided in~\cite{PEresult_GWTC2} and~\cite{ligo_scientific_collaboration_and_virgo_2021_5117703} for events in the GWTC-2 and GWTC-2.1 catalogs respectively, while the \texttt{C01:Mixed} for the O3b events newly reported in the GWTC-3 dataset~\cite{ligo_scientific_collaboration_and_virgo_2021_5546663}.
Notice that all the events, apart from those released in the GWTC-1 catalog, were analysed by the LVKC with waveform models including the effects of spin precession and higher-order modes.

In this study, we fit the BBH spin distribution by focusing only on $\chieff$. Therefore, 
we implicitly assume that the remaining spin
degrees of freedom follow the priors adopted in the  parameter estimation
(uniform in magnitude and orientation), conditioned on the proposed
$\chieff$ distribution. 
We expect this assumption to have a negligible
effect as the other spin degrees of freedom are only poorly measured.

\bibliography{main}

\end{document}